\documentclass[12pt]{article}

\usepackage{amsmath,amssymb,amsthm}
\usepackage{algorithm}
\usepackage{ascmac}
\usepackage{bm}
\usepackage{booktabs}
\usepackage{caption}
\usepackage{color}
\usepackage{colortbl}
\usepackage{comment}
\usepackage{diagbox}
\usepackage{enumerate}
\usepackage[margin=2.2cm]{geometry}
\usepackage{graphicx}
\usepackage{here}
\usepackage{latexsym}
\usepackage{lineno}
\usepackage{mathabx}
\usepackage{mathrsfs}
\usepackage{mathtools}
\mathtoolsset{showonlyrefs=true}
\usepackage{mdframed}
\usepackage{natbib}
\usepackage{setspace}
\usepackage{subcaption}
\usepackage{tikz}
\usetikzlibrary{arrows.meta,fit}

\DeclareMathAlphabet{\lmathcal}{U}{BOONDOX-calo}{m}{n}

\theoremstyle{definition}
\newtheorem{theorem}{Theorem}[section]
\newtheorem{proposition}{Proposition}[section]

\newtheorem{assumption}{Assumption}[section]
\newtheorem{definition}{Definition}[section]

\newtheorem{example}{Example}[section]
\newtheorem{remark}{Remark}[section]

\DeclareMathOperator*{\obs}{\mathrm{obs}}

\renewcommand{\hat}{\widehat}
\renewcommand{\tilde}{\widetilde}

\definecolor{mycolor}{cmyk}{0.82,0.23,0.26,0}

\usepackage[
    colorlinks=true,
    linkcolor=mycolor,
    citecolor=mycolor,
    urlcolor=mycolor,
    filecolor=mycolor,
    backref=page
]{hyperref}

\title{Selective Randomization Inference for Interference}

\author{
    Takahide Yanagi\thanks{Graduate School of Economics, Kyoto University, Yoshida Honmachi, Sakyo, Kyoto, 606-8501, Japan. \\
    Email: \href{mailto:yanagi@econ.kyoto-u.ac.jp}{yanagi@econ.kyoto-u.ac.jp}.}
}

\begin{document}

\maketitle
	
\begin{abstract}
    Empirical researchers often use tests for interference as pretests or robustness checks.
    However, when the results guide the choice of models, parameters, or hypotheses, the resulting data-dependent selection can compromise the validity of subsequent inference.
    This paper develops a two-stage selective randomization inference framework for interference that explicitly conditions on the selection event to account for data-dependent selection.
    We first test the null hypothesis of no treatment spillover using a conditional randomization method.
    If this null is rejected, we test the null hypothesis of zero exposure contrast under a pre-specified exposure mapping that allows for spillover effects;
    otherwise, we test the null hypothesis of no direct effects under the assumption of no spillover.
    We prove that each second-stage test controls the selective Type I error rate in finite samples and evaluate the size and power of the proposed tests through Monte Carlo experiments.

	\bigskip
	
\end{abstract}

\clearpage

\section{Introduction} \label{sec:introduction}

Knowledge of interactions between units is important for empirical research in the social sciences.
In particular, understanding whether and to what extent treatment of one unit affects the outcomes of other units (i.e., the existence and extent of \textit{interference}) can guide identification and statistical inference, and inform the interpretation of empirical results.
Motivated by these considerations, the recent methodological literature has devoted considerable attention to testing hypotheses concerning interference, with the null hypothesis of no treatment spillover being a representative example (\citealp{aronow2012general,athey2018exact,basse2019randomization,puelz2022graph}).
See \citet{zhang2023randomization} for a review of this literature.

In line with this, empirical researchers often test for interference either before proceeding with the main empirical analysis or after conducting it.
If researchers find evidence of interference, this motivates statistical analysis that explicitly allows for spillover effects; otherwise, researchers may focus on the direct effect under the assumption of no treatment spillover.\footnote{
    For example, \citet{bennett2021better} use randomization tests to detect spillover effects of a parent information intervention on student attendance.
    They find statistically significant spillover effects and incorporate them into their analysis of treatment targeting.
    For another example, \citet{imai2020nonpartisan} use randomization tests to assess whether treatment spillover could explain the absence of detectable total treatment effects of two social programs on voter turnout and incumbent support.
    Their tests do not reject the null hypothesis of no spillover, which may motivate interpreting the estimated total treatment effects as direct effects.
}
When the results of pretests or robustness checks guide the choice of models, parameters, or hypotheses, the resulting analysis involves data-dependent selection.
Such data-dependent selection, however, can lead to biased estimates for the selected parameters and distortions in testing the selected hypotheses under the selected model specification unless this selection is appropriately accounted for, as highlighted in the literature on post-selection and selective inference (\citealp{kuchibhotla2022post}; \citealp{neufeld2026inference}).

To account for such data-dependent selection, this paper develops a selective randomization inference framework for interference.
We consider the following two-stage randomized experiment for a population partitioned into disjoint clusters, often referred to as the randomized saturation design (\citealp{hudgens2008toward,baird2018optimal,basse2019randomization,liu2026randomization}):
the first-stage randomization assigns saturation labels to clusters to specify a target proportion of treated units within each cluster;
the second-stage randomization assigns treatments to units within clusters conditional on the assigned saturation labels.
While our selective randomization inference can be extended to other network structures and experimental designs, we focus on the randomized saturation design because the exposure mapping can be naturally defined in terms of a unit's own treatment assignment and its cluster's saturation assignment and because of its empirical relevance.\footnote{
    Many empirical studies examine both direct and spillover effects using randomized saturation designs in a variety of contexts, including cash transfer programs (e.g., \citealp{baird2011cash}), school voucher programs (e.g., \citealp{muralidharan2015aggregate}), job placement assistance (e.g., \citealp{crepon2013labor}), and loan product experiments (e.g., \citealp{cai2024indirect}).
}

The proposed testing procedure has two stages of inference with the second-stage model and hypothesis determined by the first-stage test result.
We begin by testing the null hypothesis of no spillover, stating that the outcome of each unit depends only on its own treatment assignment, using a conditional randomization approach.
If the first-stage conditional randomization test (CRT) rejects the null hypothesis of no spillover, we use the pre-specified exposure mapping defined by the unit's own treatment assignment and its cluster's saturation assignment to capture both direct and spillover effects.
Following the first-stage rejection, we assume that this exposure mapping is correctly specified and test the null hypothesis of zero exposure contrast, under which each unit's potential outcomes are equal at two pre-specified exposure values, using a selective randomization approach.
In contrast, if the first-stage CRT does not reject the null hypothesis of no spillover, we maintain the assumption of no spillover in line with the first-stage non-rejection.
We then test the null hypothesis of no direct effects, stating that the unit's own outcome does not depend on the unit's own treatment assignment, using a selective randomization approach.

Two main challenges arise in conducting the first-stage CRT and the second-stage selective randomization tests (SRTs).
First, the null hypotheses of no spillover and zero exposure contrast are generally not \textit{sharp}.
Specifically, these null hypotheses do not determine the values of all potential outcomes from the observed data, which prevents direct application of the canonical Fisher randomization test (FRT).
To conduct conditional or selective randomization inference, we need to carefully examine the imputability of potential outcomes under each of these null hypotheses.
Second and more importantly, selective randomization inference requires us to determine which counterfactual treatment assignments would reproduce the observed first-stage test result.
This task is particularly challenging after the first-stage rejection because we must reconstruct counterfactual first-stage test results under the second-stage null hypothesis, without assuming that the first-stage null hypothesis holds.

We address these challenges by characterizing the sets of treatment assignments under which the potential outcomes of selected subpopulations, called \textit{focal subpopulations}, are imputable under the relevant null hypothesis, while ensuring that the counterfactual first-stage test results can be reconstructed under the selected second-stage null hypothesis.
At the first stage, following existing conditional randomization methods, we test the null hypothesis of no spillover by holding the treatment assignments of the first-stage focal subpopulation fixed to ensure the imputability of their potential outcomes.
At the second stage, however, the imputability of potential outcomes for the second-stage focal subpopulation alone does not generally suffice.
This is because reconstructing the counterfactual first-stage test results also requires imputing the potential outcomes for the first-stage focal subpopulation under the selected second-stage null hypothesis.
In addition, the reconstruction must account for possible changes in the first-stage focal subpopulation and its conditional randomization distribution across counterfactual assignments.
The key innovation of this paper lies in jointly specifying the model assumptions, focal subpopulations, and corresponding sets of treatment assignments so that both the first-stage selection event and the potential outcomes for the second-stage focal subpopulation can be reconstructed under the selected second-stage null hypothesis.

We examine the finite-sample properties of the proposed CRT and SRTs through theoretical analysis and numerical exercises.
We prove that the first-stage CRT controls the conditional Type I error rate given the selected focal subpopulation and the restricted assignment set.
We then show that each second-stage SRT controls the selective Type I error rate given the first-stage selection event, the selected focal subpopulations, and the corresponding restricted assignment set.
We demonstrate through extensive Monte Carlo experiments that the proposed tests lead to empirical Type I error rates close to the nominal levels under the relevant null hypotheses and have adequate power under various alternatives.
We also illustrate the proposed tests using data from the cash transfer experiment of \citet{baird2011cash} by conducting selective randomization inference for direct effects conditional on the first-stage non-rejection.

This paper advances the literature on tests for interference.
Many studies develop CRTs (\citealp{aronow2012general}; \citealp{athey2018exact}; \citealp{basse2019randomization}; \citealp{puelz2022graph}; \citealp{hoshino2026conditional}; \citealp{liu2026randomization}), while recent work also explores alternative approaches (\citealp{zhong2025unconditional}; \citealp{kaido2025testing}).
To the best of our knowledge, this is the first study to develop selective randomization inference procedures that explicitly condition on the result of a test for interference.
Moreover, we provide new insights into conditional randomization in randomized saturation designs, including a tractable construction of focal subpopulations and an explicit characterization of the conditional randomization distribution.

This paper also builds on recent work on selective randomization inference.
\citet{freidling2026selective} develop a selective randomization inference framework for adaptive experiments, where experimental designs and hypotheses may depend on previously observed data.
\citet{gao2025selective} develops SRTs for treatment effects in subgroups defined by data-dependent cutoffs of continuous biomarkers.
We extend this line of research to inference following a pretest for interference by constructing focal subpopulations and sets of treatment assignments that preserve the first-stage conditioning events and permit reconstruction of the counterfactual first-stage test results under the selected second-stage null hypothesis.

The rest of the paper is organized as follows.
Section \ref{sec:setup} introduces the setup.
Section \ref{sec:roadmap} provides a roadmap to our selective randomization inference.
Sections \ref{sec:first-stage} and \ref{sec:second-stage} formalize the first-stage CRT and the second-stage SRTs, respectively.
Section \ref{sec:simulation} reports the Monte Carlo simulation results.
Section \ref{sec:application} presents the empirical illustration.
The appendices provide proofs, computational details, additional simulation results, and algorithms for the proposed tests.

\section{Setup} \label{sec:setup}

\subsection{Finite-population framework}

We adopt a finite-population perspective in which the sample at hand coincides with the population of interest.
Consider a finite population of $N$ units.
Each unit is indexed by $i \in [N]$, where we denote $[m] \coloneqq \{ 1, 2, \dots, m \}$ for any positive integer $m$.

The population is partitioned into $C$ disjoint clusters, denoted by $\mathcal{C}_1, \dots, \mathcal{C}_C$, with $\bigcup_{c=1}^C \mathcal{C}_c = [N]$.
Let $N_c \coloneqq |\mathcal{C}_c|$ denote the size of cluster $c$, which may vary across clusters.
Let $c(i) \in [C]$ denote the index of the cluster containing unit $i$, that is, $i \in \mathcal{C}_{c(i)}$.

\subsection{Randomized saturation design}

The researcher randomly assigns the binary treatment vector $Z \coloneqq (Z_i: i \in [N])$ to the population, where $Z_i \in \{ 0, 1 \}$ indicates whether unit $i$ receives treatment.
For this purpose, the researcher a priori specifies the treatment assignment mechanism.

Throughout the paper, we focus on the randomized saturation design, which consists of the following two stages of treatment assignment (\citealp{hudgens2008toward}; \citealp{liu2014large}; \citealp{baird2018optimal}; \citealp{basse2018analyzing}; \citealp{basse2019randomization}; \citealp{liu2026randomization}).
In the first stage of treatment assignment, clusters are assigned saturation labels by complete randomization.
Let $\mathcal{S} \coloneqq \{ s_1, \dots, s_L \} \subset [0, 1]$ be a pre-specified set of $L \ge 2$ distinct saturation labels, each representing a target proportion of treated units within a cluster.
The set $\mathcal{S}$ may include either, both, or neither of zero and one.
Let $\mathcal{S}_{\mathrm{int}} \coloneqq \mathcal{S} \cap (0,1)$ denote the set of interior saturation labels, which we assume to be nonempty.
For each $c \in [C]$ and saturation label $s \in \mathcal{S}_{\mathrm{int}}$, let $N_{1c}(s) \in \{ 1, \dots, N_c - 1 \}$ be the pre-specified number of units to receive treatment in cluster $c$ when it is assigned $s$.
For example, one may set $N_{1c}(s) = \lfloor s N_c \rfloor$ or $N_{1c}(s) = \lceil s N_c \rceil$, where $\lfloor \cdot \rfloor$ and $\lceil \cdot \rceil$ denote the floor and ceiling functions, respectively.
When $0 \in \mathcal{S}$, we set $N_{1c}(0) = 0$ for all $c \in [C]$.
When $1 \in \mathcal{S}$, we set $N_{1c}(1) = N_c$ for all $c \in [C]$.
For each $\ell \in [L]$, let $C_{\ell}$ be a pre-specified positive integer denoting the number of clusters assigned label $s_{\ell}$, with $\sum_{\ell=1}^L C_{\ell} = C$.
Denote the saturation label assigned to cluster $c$ by $A_c \in \mathcal{S}$.
Then, the vector of saturation assignments $A \coloneqq (A_c : c \in [C])$ is drawn uniformly from
\begin{align*}
    \mathcal{A}
    \coloneqq \left\{ a \in \mathcal{S}^C: \sum_{c = 1}^C \bm{1}\{ a_c = s_{\ell} \} = C_{\ell} \text{ for all $\ell \in [L]$} \right\},
\end{align*}
that is, $\Pr(A = a) = |\mathcal{A}|^{-1}$ for all $a \in \mathcal{A}$.
Here, the number of possible saturation assignments is given by the following multinomial coefficient:
\begin{align*}
    |\mathcal{A}|
    = \binom{C}{C_1, C_2, \dots, C_L}
    = \frac{C!}{C_1! C_2! \cdots C_L!}.
\end{align*}

In the second stage of treatment assignment, units are assigned treatment by complete randomization within each cluster, independently across clusters, conditional on $A$.
Specifically, in each cluster $c$, $N_{1c}(A_c)$ units are selected uniformly from the $N_c$ units to receive treatment.
Combining the within-cluster assignment vectors $Z_{\mathcal{C}_c} \coloneqq (Z_i : i \in \mathcal{C}_c)$ yields the full assignment vector $Z$.
The conditional probability mass function of $Z$ given $A = a$ is given by
\begin{align*}
    \Pr(Z = z \mid A = a)
    = \prod_{c = 1}^{C} \binom{N_c}{N_{1c}(a_c)}^{-1}
\end{align*}
for all $a \in \mathcal{A}$ and all $z \in \mathrm{supp}(Z \mid A = a)$, where
\begin{align*}
    \mathrm{supp}(Z \mid A = a)
    \coloneqq \left\{ z \in \{ 0, 1 \}^N : \sum_{i \in \mathcal{C}_c} z_i = N_{1c}(a_c) \text{ for all $c \in [C]$} \right\}.
\end{align*}
For any $z$ outside this conditional support, $\Pr( Z = z \mid A = a ) = 0$.

The treatment assignment mechanism is characterized by the joint probability mass function $\Pr(A = a, Z = z) = \Pr(Z = z \mid A = a) \Pr(A = a)$.
Let $A^{\obs} \coloneqq (A_c^{\obs}: c \in [C])$ and $Z^{\obs} \coloneqq (Z_i^{\obs}: i \in [N])$ denote the observed vectors of saturation assignments and treatment assignments, respectively.
For randomization inference, we use $(A, Z)$ to denote a draw from the treatment assignment mechanism independent of $(A^{\obs}, Z^{\obs})$, which itself follows the same treatment assignment mechanism.
The marginal support of $Z$ is given by
\begin{align*}
    \mathcal{Z} 
    \coloneqq \bigcup_{a \in \mathcal{A}} \mathrm{supp}(Z \mid A = a).
\end{align*}

Throughout the paper, we assume that distinct saturation labels correspond to distinct numbers of treated units.
Specifically, we assume that $N_{1c}(s) \neq N_{1c}(s')$ for all $c \in [C]$ and all $s, s' \in \mathcal{S}$ such that $s \neq s'$.
This assumption ensures that each cluster's saturation assignment can be recovered from the number of treated units in the cluster.
It therefore allows us to express the exposure mapping in the next section as a function of $Z$ alone.

\subsection{Potential outcome framework}

Suppose that the potential outcome of unit $i$ under each treatment assignment $z \in \mathcal{Z}$ is well defined, denoted by $Y_i(z)$.
The observed outcome is given by $Y_i^{\obs} \coloneqq Y_i(Z^{\obs})$.
Importantly, this notation allows unit $i$'s outcome to depend on the treatment assignments of other units as well as its own treatment assignment.
For any subpopulation $\mathcal{F} \subseteq [N]$, let $Y_{\mathcal{F}}(z) \coloneqq ( Y_i(z): i \in \mathcal{F} )$ and $Y_{\mathcal{F}}^{\obs} \coloneqq ( Y_i^{\obs}: i \in \mathcal{F} )$ denote its potential and observed outcome vectors, respectively.
In particular, we write $Y(z) \coloneqq ( Y_i(z): i \in [N] )$ and $Y^{\obs} \coloneqq (Y_i^{\obs} : i \in [N])$ for the entire population $[N]$.
Let $W \coloneqq \{ Y(z): z \in \mathcal{Z} \}$ denote the potential outcome schedule.

In line with the literature on randomization inference, we adopt a design-based inference approach.
Specifically, the treatment assignment mechanism is the only source of randomness, and the potential outcome schedule $W$ and the clusters $\mathcal{C}_1, \dots, \mathcal{C}_C$ are treated as fixed.

\section{Roadmap to Selective Randomization Inference} \label{sec:roadmap}

In the first stage, we test the null hypothesis of no treatment spillover:
\begin{align} \label{eq:H1}
    H_1: \qquad
    \text{$Y_i(z) = Y_i(z')$ for all $i \in [N]$ and all $z, z' \in \mathcal{Z}$ such that $z_i = z_i'$}.
\end{align}
This null hypothesis has been studied in the literature on randomization inference for interference; see, for example, \citet[Hypothesis 2]{athey2018exact} and \citet[Section 6]{zhang2025multiple}.
If $H_1$ is true, the outcome of each unit depends only on its own treatment, and the potential outcome given $Z_i = z_i$ is thus well defined, which is denoted by $\tilde{y}_i(z_i)$.
Rejection of $H_1$ provides statistical evidence for the existence of treatment spillover.
In Section \ref{sec:first-stage}, we develop a conditional randomization approach for testing $H_1$.

If the first-stage CRT rejects $H_1$, we next conduct selective randomization inference for direct or spillover effects based on a pre-specified exposure mapping $E_i$.
Following the literature on within-cluster interference, we summarize each unit's exposure by its own treatment assignment and the total number of treated units in its cluster.
Since we assume that distinct saturation labels correspond to distinct numbers of treated units, there is a one-to-one correspondence between the total number of treated units and the saturation label for each cluster.
Based on this, we consider the following form of exposure mapping:
\begin{align} \label{eq:exposure}
    E_i(Z)
    \coloneqq (Z_i, A_{c(i)}).
\end{align}
We say that the exposure mapping $E_i$ is correct if the following condition holds:
\begin{align} \label{eq:correct-exposure}
    \text{For all $i \in [N]$ and $z, z' \in \mathcal{Z}$, $Y_i(z) = Y_i(z')$ whenever $E_i(z) = E_i(z')$}.
\end{align}
If this is true, the potential outcome $Y_i(Z)$ depends on the treatment assignments $Z$ only through the unit's own treatment assignment $Z_i$ and the saturation assignment $A_{c(i)}$.
Then, letting $\mathcal{E}_i \coloneqq \{ E_i(z): z \in \mathcal{Z} \}$, there exists a function $y_i: \mathcal{E}_i \to \mathbb{R}$ such that $Y_i(z) = y_i(E_i(z))$ for all $z \in \mathcal{Z}$.
In other words, $y_i(e)$ is the potential outcome given $E_i = e$ for $e \in \mathcal{E}_i$, where we denote $E_i \coloneqq E_i(Z)$ with an abuse of notation.

After the rejection of the null hypothesis of no spillover $H_1$, we proceed under the assumption that the pre-specified exposure mapping $E_i$ is correct in the sense of \eqref{eq:correct-exposure}.
Under this assumption, we test the null hypothesis of zero exposure contrast between two pre-specified distinct exposure values $\mathtt{a}, \mathtt{b} \in \bigcap_{i \in [N]} \mathcal{E}_i$ with $\mathtt{a} \neq \mathtt{b}$:
\begin{align} \label{eq:H2}
    H_2: \qquad
    y_i(\mathtt{a}) = y_i(\mathtt{b}) \quad \text{for all } i \in [N].
\end{align}
In the literature on randomization inference for interference, \citet{basse2019randomization}, \citet{puelz2022graph}, and \citet{liu2026randomization} have studied similar null hypotheses for testing direct and spillover effects.
For example, if $\mathtt{a} = (1, s)$ and $\mathtt{b} = (0, s)$ for $s \in \mathcal{S}_{\mathrm{int}}$, then $H_2$ states that there are no direct effects at this interior saturation label $s$, where the direct effect is defined as $y_i(1, s) - y_i(0, s)$.
Another example takes $\mathtt{a} = (0, s)$ and $\mathtt{b} = (0, s')$ for distinct $s, s' \in \mathcal{S} \setminus \{ 1 \}$.
In this case, $H_2$ states that any unit has the same potential outcome under these two saturation labels when assigned to control.
Under the assumption that the exposure mapping is correctly specified, rejection of $H_2$ provides statistical evidence for a nonzero exposure contrast for at least one unit.
Section \ref{subsec:reject} develops an SRT for $H_2$ conditional on the rejection of $H_1$.
Moreover, if we additionally assume that the exposure effect $y_i(\mathtt{a}) - y_i(\mathtt{b})$ is constant across units, we can construct a selective confidence set for the average exposure effect $N^{-1} \sum_{i=1}^N [ y_i(\mathtt{a}) - y_i(\mathtt{b}) ]$ by test inversion.
See Remark \ref{remark:CI1} for details.

If the first-stage CRT does not reject $H_1$, we assume no treatment spillover in the sense of \eqref{eq:H1} and test the null hypothesis of no direct effects:
\begin{align} \label{eq:tildeH1}
    \tilde{H}_1: \qquad
    \tilde{y}_i(1) = \tilde{y}_i(0) \quad \text{for all } i \in [N],
\end{align}
where $\tilde{y}_i(z_i)$ denotes the potential outcome given $Z_i = z_i$, as defined above.
Note that the combination of $H_1$ and $\tilde{H}_1$ is identical to Fisher's sharp null stating that $Y_i(z) = Y_i(z')$ for all $i \in [N]$ and $z, z' \in \mathcal{Z}$.
In the absence of treatment spillover, rejection of $\tilde{H}_1$ provides statistical evidence of a nonzero direct effect for at least one unit, where the direct effect is defined as $\tilde{y}_i(1) - \tilde{y}_i(0)$.
Because $\tilde{H}_1$ is selected based on the non-rejection of $H_1$, Section \ref{subsec:accept} develops an SRT that conditions on this selection event.
As in the first-stage rejection case, we can also construct a selective confidence set for the average direct effect $N^{-1} \sum_{i=1}^N [ \tilde y_i(1) - \tilde y_i(0) ]$ by test inversion, assuming that $\tilde y_i(1) - \tilde y_i(0)$ is constant across units.

\begin{remark}
    Even when the first-stage CRT does not reject $H_1$, researchers may wish to conduct inference on direct effects while allowing for treatment spillover.
    Specifically, researchers may test $H_2$ with $\mathtt{a} = (1, s)$ and $\mathtt{b} = (0, s)$ for a pre-specified interior saturation label $s \in \mathcal{S}_{\mathrm{int}}$, rather than $\tilde{H}_1$.
    Assuming that the exposure mapping $E_i$ is correct, this null hypothesis rules out direct effects at saturation label $s$ while allowing for spillover effects.
    Importantly, a CRT for $H_2$ that does not account for the first-stage selection need not control the selective Type I error rate.
    To obtain inference that is valid conditional on the first-stage non-rejection, researchers should adapt the selective randomization inference procedure developed in Section \ref{subsec:reject} by conditioning on non-rejection rather than rejection.
\end{remark}

\section{First Stage: Conditional Randomization Test} \label{sec:first-stage}

In this section, we develop a CRT for the null hypothesis of no spillover $H_1$ defined in \eqref{eq:H1}.
We characterize the potential outcomes that are imputable under $H_1$, formalize the testing procedure, and discuss the choice of focal subpopulation.

\subsection{Imputability of potential outcomes} \label{subsec:first-impute}

A key challenge in testing $H_1$ is that only a part of the potential outcomes are imputable, unlike Fisher's sharp null under which all potential outcomes are imputable.
The next definitions formalize imputability of potential outcomes, sharp null hypotheses, and partially sharp null hypotheses, following \citet[Definition 1]{zhang2023randomization} and \citet[Definition 1]{zhong2025unconditional}.

\begin{definition}[Imputability] \label{def:imputable}
    Let $H$ be a null hypothesis that restricts the potential outcome schedule $W$.
    \begin{enumerate}[(i)]
        \item For a given $i \in [N]$ and a given $z \in \mathcal{Z}$, the potential outcome $Y_i(z)$ is \textit{imputable with respect to $H$} if it is uniquely determined by the observed data under $H$. 
        \item For a given $i \in [N]$, a set $\mathcal{Z}_i \subseteq \mathcal{Z}$ is an \textit{imputable set} of unit $i$ with respect to $H$ if $Y_i(z)$ is imputable with respect to $H$ for all $z \in \mathcal{Z}_i$.
    \end{enumerate}
\end{definition}

\begin{definition}[Sharp null hypothesis] \label{def:sharp}
    A null hypothesis $H$ is \textit{sharp} if $Y_i(z)$ is imputable with respect to $H$ for all $i \in [N]$ and all $z \in \mathcal{Z}$, that is, $\mathcal{Z}_i = \mathcal{Z}$ is the largest imputable set of unit $i$ with respect to $H$ for all $i \in [N]$.
\end{definition}

\begin{definition}[Partially sharp null hypothesis] \label{def:partial}
    Given the observed data, a null hypothesis $H$ is \textit{partially sharp} if it is not sharp but there exists at least one pair $(i, z)$ with $i \in [N]$ and $z \in \mathcal{Z} \setminus \{ Z^{\obs} \}$ such that $Y_i(z)$ is imputable with respect to $H$.
\end{definition}

Based on these definitions, we examine the partial sharpness of the null hypothesis of no spillover $H_1$.
Take an arbitrary unit $i \in [N]$.
For any $z \in \mathcal{Z}$ such that $z_i = Z_i^{\obs}$, the potential outcome $Y_i(z)$ is imputable with respect to $H_1$ because $Y_i(z) = Y_i^{\obs}$ under $H_1$.
In contrast, for any $z' \in \mathcal{Z}$ such that $z'_i \neq Z_i^{\obs}$, the observed data do not determine the value of $Y_i(z')$ under $H_1$, and hence $Y_i(z')$ is not imputable with respect to $H_1$.
Thus, the largest imputable set of unit $i \in [N]$ with respect to $H_1$ is given by
\begin{align} \label{eq:mathcal-Z-1}
    \mathcal{Z}_i^1 
    \coloneqq \left\{ z \in \mathcal{Z}: z_i = Z_i^{\obs} \right\}.
\end{align}
In general, $\mathcal{Z}_i^1$ contains many assignments other than $Z^{\obs}$, so that $H_1$ is partially sharp.

A key challenge in testing $H_1$ is that the potential outcome vector $Y(z)$ on the entire sample $[N]$ is not imputable for any $z \in \mathcal{Z} \setminus \{ Z^{\obs} \}$, which precludes implementation of the canonical FRT that requires the imputation of all potential outcome vectors.
Indeed, the intersection of imputable sets $\bigcap_{i \in [N]} \mathcal{Z}_i^1$ over the entire sample $[N]$, which is the set of assignments $z$ for which $Y(z)$ on $[N]$ is imputable with respect to $H_1$, reduces to $\{ Z^{\obs} \}$.
Accordingly, the only imputable potential outcome vector on $[N]$ is $Y^{\obs} = Y(Z^{\obs})$.

The following example further helps to understand the issue of imputability.

\begin{example}[Six units in two clusters] \label{example1}
    Suppose that there are two disjoint clusters and each cluster $c \in \{ 1, 2 \}$ includes three units as in Figure \ref{fig:cluster1}, that is, $N = 6$, $C = 2$, and $N_c = 3$ for each $c \in \{ 1, 2 \}$.
    Although real applications involve more units and clusters, we use this simple example to illustrate the main ideas.
    The researcher adopts the randomized saturation design with the set of saturation labels $\mathcal{S} = \{ s_1, s_2 \} = \{ 1/3, 2/3 \}$, corresponding to $N_{1c}(1/3) = 1$ and $N_{1c}(2/3) = 2$ for each $c$.
    One cluster is assigned label $s_1 = 1/3$ and the other cluster is assigned label $s_2 = 2/3$, so that $C_1 = C_2 = 1$.
    In this setting, the support $\mathcal{Z}$ consists of $2 \times \binom{3}{1} \times \binom{3}{2} = 18$ elements, listed in Table \ref{tab:example1}.
    Suppose that the observed saturation assignments are $A_1^{\obs} = 1/3$ and $A_2^{\obs} = 2/3$.
    Let $Z^{\obs} = (1, 0, 0, 1, 1, 0)$ and $Y^{\obs} = (11, 12, 13, 14, 15, 16)$ be the observed treatment and outcome vectors, respectively.
    As can be seen from Table \ref{tab:example1}, all units have nine potential outcomes that are not imputable with respect to the null hypothesis of no spillover $H_1$, implying that $H_1$ is not sharp.
    Indeed, the table shows that the only imputable potential outcome vector on $[N]$ is $Y^{\obs} = Y(Z^{\obs})$.
    However, the largest imputable set $\mathcal{Z}_i^1$ for each unit contains nine assignments, including eight other than $Z^{\obs}$.
    For example, for unit 1, $\mathcal{Z}_1^1 = \{ z_{(1)}, \dots, z_{(3)}, z_{(10)}, \dots, z_{(15)} \}$.
    Thus, $H_1$ is partially sharp.
\end{example}

\begin{figure}[p]
    \centering
    \begin{tikzpicture}[
        person/.style={circle, draw, minimum size=10mm},
        cluster/.style={
            draw,
            rounded corners,
            thick,
            inner sep=8mm,
            label={[font=\bfseries]above:#1}
        }
    ]

        \node[person,fill=cyan!30] (1) at (0,0) {$1$};
        \node[person] (2) at (1.2,0) {$2$};
        \node[person] (3) at (2.4,0) {$3$};

        \node[cluster=Cluster 1, fit=(1) (2) (3)] {};

        \node[person,fill=cyan!30] (4) at (6,0) {$4$};
        \node[person,fill=cyan!30] (5) at (7.2,0) {$5$};
        \node[person] (6) at (8.4,0) {$6$};

        \node[cluster=Cluster 2, fit=(4) (5) (6)] {};

    \end{tikzpicture}
    \caption{Cluster structure and the observed assignment $Z^{\obs}$ in Example \ref{example1}. 
    Cyan and white nodes represent treated and control units, respectively.} \label{fig:cluster1}

    \par\vspace{2cm}

    \begin{minipage}{\textwidth}
    \captionof{table}{Treatment assignments $z \in \mathcal{Z}$ and imputable potential outcomes $Y_i(z)$ under the null hypothesis of no spillover $H_1$ in Example \ref{example1}. 
    Red question marks indicate potential outcomes that are not imputable under $H_1$.
    The columns labeled ``\textbf{Focal 1}'' indicate the focal units $\mathcal{F}_1^{\obs} = \{ 3, 4 \}$ used in Example \ref{example2}, and shaded rows indicate treatment assignments in the largest imputable set $\mathcal{Z}_{\mathcal{F}_1^{\obs}}^1 = \{ Z^{\obs}, z_{(2)}, z_{(4)}, z_{(5)}, z_{(10)} \}$.}
    \label{tab:example1}
    \centering
    \begin{tabular}{r|cccccc}
         & \multicolumn{6}{c}{The potential outcome $Y_i(z)$} \\
        \diagbox[width=15em]{Assignment $z$}{Unit $i$} & $i = 1$ & $i = 2$ & $\bm{i = 3}$ & $\bm{i = 4}$ & $i = 5$ & $i = 6$ \\
        & & & {\scriptsize \textbf{Focal 1}} & {\scriptsize \textbf{Focal 1}} & & \\
        \hline
        \multicolumn{1}{l|}{Given $A_1 = 1/3$ and $A_2 = 2/3$:} \\
        \rowcolor{gray!30} $Z^{\obs} = z_{(1)} = (1, 0, 0, 1, 1, 0)$ & 11 & 12 & 13 & 14 & 15 & 16 \\
        \rowcolor{gray!30} $z_{(2)} = (1, 0, 0, 1, 0, 1)$ & 11 & 12 & 13 & 14 & \textcolor{red}{?} & \textcolor{red}{?} \\
        $z_{(3)} = (1, 0, 0, 0, 1, 1)$ & 11 & 12 & 13 & \textcolor{red}{?} & 15 & \textcolor{red}{?} \\
        \rowcolor{gray!30} $z_{(4)} = (0, 1, 0, 1, 1, 0)$ & \textcolor{red}{?} & \textcolor{red}{?} & 13 & 14 & 15 & 16 \\
        \rowcolor{gray!30} $z_{(5)} = (0, 1, 0, 1, 0, 1)$ & \textcolor{red}{?} & \textcolor{red}{?} & 13 & 14 & \textcolor{red}{?} & \textcolor{red}{?} \\
        $z_{(6)} = (0, 1, 0, 0, 1, 1)$ & \textcolor{red}{?} & \textcolor{red}{?} & 13 & \textcolor{red}{?} & 15 & \textcolor{red}{?} \\
        $z_{(7)} = (0, 0, 1, 1, 1, 0)$ & \textcolor{red}{?} & 12 & \textcolor{red}{?} & 14 & 15 & 16 \\
        $z_{(8)} = (0, 0, 1, 1, 0, 1)$ & \textcolor{red}{?} & 12 & \textcolor{red}{?} & 14 & \textcolor{red}{?} & \textcolor{red}{?} \\
        $z_{(9)} = (0, 0, 1, 0, 1, 1)$ & \textcolor{red}{?} & 12 & \textcolor{red}{?} & \textcolor{red}{?} & 15 & \textcolor{red}{?} \\
        \\
        \multicolumn{1}{l|}{Given $A_1 = 2/3$ and $A_2 = 1/3$:} \\
        \rowcolor{gray!30} $z_{(10)} = (1, 1, 0, 1, 0, 0)$ & 11 & \textcolor{red}{?} & 13 & 14 & \textcolor{red}{?} & 16 \\
        $z_{(11)} = (1, 1, 0, 0, 1, 0)$ & 11 & \textcolor{red}{?} & 13 & \textcolor{red}{?} & 15 & 16 \\
        $z_{(12)} = (1, 1, 0, 0, 0, 1)$ & 11 & \textcolor{red}{?} & 13 & \textcolor{red}{?} & \textcolor{red}{?} & \textcolor{red}{?} \\
        $z_{(13)} = (1, 0, 1, 1, 0, 0)$ & 11 & 12 & \textcolor{red}{?} & 14 & \textcolor{red}{?} & 16 \\
        $z_{(14)} = (1, 0, 1, 0, 1, 0)$ & 11 & 12 & \textcolor{red}{?} & \textcolor{red}{?} & 15 & 16 \\
        $z_{(15)} = (1, 0, 1, 0, 0, 1)$ & 11 & 12 & \textcolor{red}{?} & \textcolor{red}{?} & \textcolor{red}{?} & \textcolor{red}{?} \\
        $z_{(16)} = (0, 1, 1, 1, 0, 0)$ & \textcolor{red}{?} & \textcolor{red}{?} & \textcolor{red}{?} & 14 & \textcolor{red}{?} & 16 \\
        $z_{(17)} = (0, 1, 1, 0, 1, 0)$ & \textcolor{red}{?} & \textcolor{red}{?} & \textcolor{red}{?} & \textcolor{red}{?} & 15 & 16 \\
        $z_{(18)} = (0, 1, 1, 0, 0, 1)$ & \textcolor{red}{?} & \textcolor{red}{?} & \textcolor{red}{?} & \textcolor{red}{?} & \textcolor{red}{?} & \textcolor{red}{?} \\
    \end{tabular}
    \end{minipage}
\end{figure}

\subsection{Conditional randomization test} \label{subsec:first-test}

To test the null hypothesis of no spillover $H_1$, we adopt a conditional randomization approach that selects a subpopulation of units and restricts randomization to assignments under which the potential outcome vector for the selected subpopulation is imputable.

Specifically, let $\mathcal{F}_1: \mathcal{Z} \to 2^{[N]}$ be a pre-specified set-valued function that assigns to each assignment $z \in \mathcal{Z}$ a nonempty proper subset of the population $[N]$, that is,
\begin{align} \label{eq:first-focal}
    \text{$\mathcal{F}_1(z) \subset [N]$ and $\mathcal{F}_1(z) \neq \varnothing$ for all $z \in \mathcal{Z}$.}
\end{align}
The set $\mathcal{F}_1(z)$ represents the \textit{focal subpopulation} that would be selected if assignment $z$ were observed, and its members are referred to as the corresponding \textit{focal units}.
Let $F_{1i}(z) \coloneqq \bm{1}\{ i \in \mathcal{F}_1(z) \}$ indicate whether unit $i$ is a focal unit under treatment assignment $z$, with the corresponding indicator vector $F_1(z) \coloneqq (F_{1i}(z): i \in [N])$.
Denote the \textit{observed} focal subpopulation by $\mathcal{F}_1^{\obs} \coloneqq \mathcal{F}_1(Z^{\obs})$, with membership indicator $F_{1i}^{\obs} \coloneqq F_{1i}(Z^{\obs})$ and indicator vector $F_1^{\obs} \coloneqq (F_{1i}^{\obs}: i \in [N])$.
Any pre-specified focal-subpopulation selection function satisfying \eqref{eq:first-focal} may be used, provided that the conditional randomization distribution accounts for it.
However, the choice of focal subpopulation, like the choice of test statistic, is important for the power of the test.
We discuss how to specify the focal subpopulation in practice in Section \ref{subsec:first-focal}.

\begin{remark}
    The focal subpopulation $\mathcal{F}_1^{\obs}$ may also depend on auxiliary randomness independent of the treatment assignment.
    Let $U^{\obs} \in \mathcal{U}$ be an observed auxiliary random vector generated independently of $Z^{\obs}$.
    For example, $U^{\obs}$ may consist of independent uniform random variables used to rank units for the selection of focal units.
    Let $\mathcal{F}_1: \mathcal{Z} \times \mathcal{U} \to 2^{[N]}$ be a pre-specified set-valued function that returns a nonempty proper subset of $[N]$.
    Define $F_{1i}(z, u) \coloneqq \bm{1}\{ i \in \mathcal{F}_1(z, u) \}$ and $F_1(z, u) \coloneqq (F_{1i}(z, u): i \in [N])$.
    The observed focal subpopulation and its indicator vector are then $\mathcal{F}_1^{\obs} = \mathcal{F}_1(Z^{\obs}, U^{\obs})$ and $F_1^{\obs} = F_1(Z^{\obs}, U^{\obs})$, respectively.
    Even in this case, the CRT developed below is still valid by holding $U^{\obs}$ fixed.
    In what follows, for notational simplicity, any auxiliary randomness used for the selection of focal units is held fixed, and its role in the selection rule and conditioning statements is suppressed in the notation.
    Thus, $\mathcal{F}_1(z)$ may be interpreted as $\mathcal{F}_1(z, U^{\obs})$, and all subsequent results should be understood conditional on $U^{\obs}$.
\end{remark}

Once the focal subpopulation $\mathcal{F}_1^{\obs}$ is selected, we can characterize the set of assignments under which the potential outcome vector on $\mathcal{F}_1^{\obs}$ is imputable.
Recall that, for each focal unit $i \in \mathcal{F}_1^{\obs}$, the largest imputable set $\mathcal{Z}_i^1$ with respect to $H_1$ is given by \eqref{eq:mathcal-Z-1}.
We can then construct the largest imputable set for $\mathcal{F}_1^{\obs}$, consisting of assignments $z$ under which the potential outcome vector $Y_{\mathcal{F}_1^{\obs}}(z)$ is imputable under $H_1$, as follows:
\begin{align} \label{eq:mathcal-I-F1}
    \mathcal{Z}_{\mathcal{F}_1^{\obs}}^1
    \coloneqq \bigcap_{i \in \mathcal{F}_1^{\obs}} \mathcal{Z}_i^1
    = \left\{ z \in \mathcal{Z}: \text{$z_i = Z_i^{\obs}$ for all $i \in \mathcal{F}_1^{\obs}$} \right\}.
\end{align}
For notational simplicity, we often write $\mathcal{Z}^1 \coloneqq \mathcal{Z}_{\mathcal{F}_1^{\obs}}^1$ when no confusion arises.
Importantly, since $\mathcal{F}_1^{\obs}$ is a proper subset of $[N]$, the imputable set $\mathcal{Z}^1$ may contain assignments beyond the observed assignment $Z^{\obs}$.
By construction, for any $z \in \mathcal{Z}^1$, the potential outcome vector $Y_{\mathcal{F}_1^{\obs}}(z)$ is imputable under $H_1$; that is, $Y_{\mathcal{F}_1^{\obs}}(z) = Y_{\mathcal{F}_1^{\obs}}^{\obs}$ under $H_1$.

Importantly, there is a tradeoff between the size of the focal subpopulation $\mathcal{F}_1^{\obs}$ and the size of the imputable set $\mathcal{Z}^1$.
Specifically, adding units to the focal subpopulation $\mathcal{F}_1^{\obs}$ can only reduce or leave unchanged the corresponding imputable set $\mathcal{Z}^1$.
To illustrate this tradeoff and highlight the role of the focal subpopulation, we continue Example \ref{example1}.

\begin{example}[Six units in two clusters, continued] \label{example2}
    In the setting of Example \ref{example1}, we can select the focal subpopulation $\mathcal{F}_1^{\obs}$ in various ways.
    If we select $\mathcal{F}_1^{\obs} = \{ 3, 4 \}$, the corresponding largest imputable set is $\mathcal{Z}^1 = \{ Z^{\obs}, z_{(2)}, z_{(4)}, z_{(5)}, z_{(10)} \}$.
    These focal units and assignments are indicated by the columns labeled ``\textbf{Focal 1}'' and the shaded rows, respectively, in Table \ref{tab:example1}.
    The imputable set $\mathcal{Z}^1$ contains five assignments, including four other than the observed assignment $Z^{\obs}$.
    Alternatively, if we add unit 1 to the focal subpopulation, so that $\mathcal{F}_1^{\obs} = \{ 1, 3, 4 \}$, the largest imputable set becomes smaller, with $\mathcal{Z}^1 = \{ Z^{\obs}, z_{(2)}, z_{(10)} \}$.
    Nevertheless, this $\mathcal{Z}^1$ still contains three assignments, including two other than $Z^{\obs}$.
\end{example}

Given the focal subpopulation $\mathcal{F}_1^{\obs}$ and the largest imputable set $\mathcal{Z}^1$, the CRT for the null hypothesis of no spillover $H_1$ proceeds as follows.
Let $T_1(a, z, Y_{\mathcal{F}_1^{\obs}}(z))$ be a generic test statistic computed with the treatment assignment $z \in \mathcal{Z}^1$, the corresponding saturation assignment $a$, and the corresponding potential outcome vector $Y_{\mathcal{F}_1^{\obs}}(z)$.
For example, we may consider the following regression model:
\begin{align} \label{eq:first-statistic}
    Y_i(z)
    = \beta_0 + z_i \beta_1 + \sum_{\ell = 1}^{L-1} \bm{1}\{ a_{c(i)} = s_{\ell + 1} \} \gamma_{\ell} + \sum_{\ell = 1}^{L-1} z_i \bm{1}\{ a_{c(i)} = s_{\ell + 1} \} \delta_{\ell} + \mathrm{error}_i
    \quad
    \text{for $i \in \mathcal{F}_1^{\obs}$}.
\end{align}
This regression model is used only to construct a test statistic, and the validity of the CRT does not require the regression model to be correctly specified.
We may use the Wald statistic for testing $\gamma_1 = \dots = \gamma_{L-1} = \delta_1 = \dots = \delta_{L-1} = 0$.
The imputability of $Y_{\mathcal{F}_1^{\obs}}(z)$ under $H_1$ allows us to impute the test statistic under $H_1$, that is, $T_1(a, z, Y_{\mathcal{F}_1^{\obs}}(z)) = T_1(a, z, Y_{\mathcal{F}_1^{\obs}}^{\obs})$ for any $z \in \mathcal{Z}^1$.
Then, the conditional $p$-value for testing $H_1$ is defined by
\begin{align} \label{eq:pval-first}
\begin{split}
    \mathrm{pval}_1(A^{\obs}, Z^{\obs})
    \coloneqq \Pr \Big[ T_1 \big( A, Z, Y_{\mathcal{F}_1^{\obs}}^{\obs} \big) \ge T_1 \big( A^{\obs}, Z^{\obs}, Y_{\mathcal{F}_1^{\obs}}^{\obs} \big) \; \Big| \; F_1(Z) = F_1^{\obs}, Z \in \mathcal{Z}^1 \Big],
\end{split}
\end{align}
where the probability is taken with respect to the conditional distribution of $(A, Z)$ induced by the randomized saturation design, given $F_1(Z) = F_1^{\obs}$ and $Z \in \mathcal{Z}^1$, with the observed data and the conditioning quantities held fixed.
For notational simplicity, we suppress the dependence of $\mathrm{pval}_1(A^{\obs}, Z^{\obs})$ on $F_1^{\obs}$ and $\mathcal{Z}^1$.
The CRT rejects $H_1$ at a significance level $\alpha_1 \in [0, 1]$ if $\mathrm{pval}_1(A^{\obs}, Z^{\obs}) \le \alpha_1$.
Algorithm \ref{alg:first-stage} in Appendix \ref{sec:algoirthm} summarizes this testing procedure.

The following proposition establishes that the CRT for $H_1$ controls the conditional Type I error rate in finite samples.

\begin{proposition} \label{prop:stage1}
    Suppose that the following three conditions hold.
    (i) Let $\mathcal{F}_1: \mathcal{Z} \to 2^{[N]}$ be a pre-specified focal-subpopulation selection function satisfying \eqref{eq:first-focal}, with the observed focal subpopulation $\mathcal{F}_1^{\obs} = \mathcal{F}_1(Z^{\obs})$ and the corresponding indicator vector $F_1^{\obs}$.
    (ii) Let $\mathcal{Z}^1 = \mathcal{Z}_{\mathcal{F}_1^{\obs}}^1$ be the largest imputable set as defined in \eqref{eq:mathcal-I-F1}.
    (iii) Let $T_1(a, z, Y_{\mathcal{F}_1^{\obs}}(z))$ be a pre-specified test statistic that is well-defined for any $z \in \mathcal{Z}^1$, the corresponding $a$, and the corresponding $Y_{\mathcal{F}_1^{\obs}}(z)$.
    Then, under $H_1$, for all $\alpha_1 \in [0, 1]$,
    \begin{align*}
        \Pr \left[ \mathrm{pval}_1(A^{\obs}, Z^{\obs}) \le \alpha_1 \;\middle|\; F_1^{\obs}, \mathcal{Z}^1 \right] \le \alpha_1
        \qquad \text{almost surely},
    \end{align*}
    where the probability is taken with respect to the conditional distribution of $(A^{\obs}, Z^{\obs})$ given $F_1^{\obs}$ and $\mathcal{Z}^1$.
\end{proposition}

We compute the conditional $p$-value based on the conditional randomization distribution:
\begin{align*}
    \pi_1(a, z)
    \coloneqq \Pr\left( A = a, Z = z \; \middle| \; F_1(Z) = F_1^{\obs}, Z \in \mathcal{Z}^1 \right).
\end{align*}
Two practical issues arise in implementing this procedure.
First, the form of $\pi_1$ generally depends on the focal-subpopulation selection function $\mathcal{F}_1$.
Second, the conditional support of $(A, Z)$ given $F_1(Z) = F_1^{\obs}$ and $Z \in \mathcal{Z}^1$ is typically large.
To address these issues, the next subsection proposes a tractable focal-subpopulation selection function, and Appendix \ref{sec:computation} derives the corresponding conditional randomization distribution $\pi_1$ and describes a Monte Carlo approximation of the conditional $p$-value.

\subsection{Choice of focal subpopulation} \label{subsec:first-focal}

As discussed above, there is a tradeoff between the number of focal units and the number of treatment assignments in conditional randomization: adding units to the focal subpopulation $\mathcal{F}_1^{\obs}$ can only reduce or leave unchanged the imputable set $\mathcal{Z}^1$.
This tradeoff does not compromise the conditional Type I error control in Proposition \ref{prop:stage1}, but it is important for the power of the test: few focal units or limited variation in the test statistic can reduce the power.

In general, selecting focal units from all available clusters should be desirable because it allows the test to use outcome information from them.
Based on this idea, we consider the following construction of the focal subpopulation.
First, we specify a nonnegative integer $K_c$ for each cluster $c$ satisfying $\sum_{c = 1}^C K_c \ge 1$ and
\begin{align} \label{eq:bound-K}
    0 \le K_c \le \min \left\{ \min_{s \in \mathcal{S}_{\mathrm{int}}} N_{1c}(s), \; N_c - \max_{s \in \mathcal{S}_{\mathrm{int}}} N_{1c}(s) \right\}.
\end{align}
Then, we select $K_c$ focal units uniformly without replacement from all $N_c$ units in each cluster $c$, independently across clusters and independently of the treatment assignment.
This procedure ensures that, with the auxiliary randomness held fixed, the focal-subpopulation selection function $\mathcal{F}_1$ is constant across treatment assignments: $\mathcal{F}_1(z) = \mathcal{F}_1^{\obs}$ for all $z \in \mathcal{Z}$.

The bound in \eqref{eq:bound-K} ensures that, within each cluster, the observed assignments of focal units are compatible with any interior saturation label, regardless of the observed saturation label.
To see this, denote the numbers of treated and control focal units in cluster $c$ by
\begin{align*}
    K_{1c} \coloneqq \sum_{i \in \mathcal{C}_c} F_{1i}^{\obs} Z_i^{\obs},
    \qquad 
    K_{0c} \coloneqq \sum_{i \in \mathcal{C}_c} F_{1i}^{\obs} (1 - Z_i^{\obs}),
\end{align*}
respectively.
By construction, $K_{1c} + K_{0c} = K_c$.
The required number of treated non-focal units in cluster $c$ under interior saturation label $s \in \mathcal{S}_{\mathrm{int}}$ is $N_{1c}(s) - K_{1c}$, which satisfies the following inequalities under \eqref{eq:bound-K}:
\begin{align*}
    N_{1c}(s) - K_{1c} &\ge \min_{s' \in \mathcal{S}_{\mathrm{int}}} N_{1c}(s') - K_c \ge 0,\\
    N_{1c}(s) - K_{1c} &\le \max_{s' \in \mathcal{S}_{\mathrm{int}}} N_{1c}(s') \le N_c - K_c.
\end{align*}
In other words, \eqref{eq:bound-K} ensures that the required number of treated non-focal units, $N_{1c}(s) - K_{1c}$, lies between zero and the total number of non-focal units, $N_c - K_c$.
Thus, any interior saturation label is compatible with the observed treatment assignments of focal units within each cluster.

If $0 \in \mathcal{S}$, the zero saturation label is compatible with the observed treatment assignments of focal units in cluster $c$ if and only if $K_{1c} = 0$, meaning that the cluster contains no treated focal units.
This condition holds automatically when $A_c^{\obs} = 0$ and may also hold when $A_c^{\obs} > 0$.
Similarly, if $1 \in \mathcal{S}$, the one saturation label is compatible with the observed treatment assignments of focal units in cluster $c$ if and only if $K_{0c} = 0$, which holds automatically when $A_c^{\obs} = 1$ and may also hold when $A_c^{\obs} < 1$.
Thus, under conditional randomization, the zero and one saturation labels can be assigned only to clusters with $K_{1c} = 0$ and $K_{0c} = 0$, respectively, as detailed in Appendix \ref{sec:computation}.

\section{Second Stage: Selective Randomization Inference} \label{sec:second-stage}

In this section, we develop the second-stage SRTs based on the result of the first-stage test.
We consider the first-stage rejection and non-rejection cases in Sections \ref{subsec:reject} and \ref{subsec:accept}, respectively.

To proceed, for any $z \in \mathcal{Z}^1$ and the corresponding saturation assignment $a$, we define
\begin{align*}
    \mathrm{pval}_1(a, z)
    \coloneqq \Pr \Big[ T_1 \big( A, Z, Y_{\mathcal{F}_1^{\obs}}(z) \big) \ge T_1 \big( a, z, Y_{\mathcal{F}_1^{\obs}}(z) \big) \; \Big| \; F_1(Z) = F_1^{\obs}, Z \in \mathcal{Z}^1 \Big],
\end{align*}
where the probability is taken with respect to the conditional distribution of $(A, Z)$ given $F_1(Z) = F_1^{\obs}$ and $Z \in \mathcal{Z}^1$, holding $(a, z)$, $Y_{\mathcal{F}_1^{\obs}}(z)$, $F_1^{\obs}$, and $\mathcal{Z}^1$ fixed.
Let
\begin{align*}
    \phi_1(a, z)
    \coloneqq \bm{1} \left\{ \mathrm{pval}_1(a, z) \le \alpha_1 \right\},
\end{align*}
so that $\phi_1^{\obs} \coloneqq \phi_1(A^{\obs}, Z^{\obs})$ indicates whether the first-stage CRT rejects the null hypothesis of no spillover $H_1$ at level $\alpha_1$.

Our selective randomization inference requires that, for $z \in \mathcal{Z}^1$ and the corresponding $a$, $\phi_1(a, z)$ can be interpreted as the counterfactual rejection indicator of the first-stage CRT if $(a, z)$ were observed instead of $(A^{\obs}, Z^{\obs})$.
However, this interpretation need not hold with a general first-stage focal subpopulation selection function $\mathcal{F}_1$ because $\phi_1(a, z)$ is defined with respect to $\mathcal{F}_1^{\obs} = \mathcal{F}_1(Z^{\obs})$ rather than $\mathcal{F}_1(z)$.
To ensure that $\phi_1(a, z)$ reproduces the counterfactual first-stage test results for all $z \in \mathcal{Z}^1$ and the corresponding $a$, we impose the following condition.

\begin{assumption} \label{as:first-focal}
    $\mathcal{F}_1(z) = \mathcal{F}_1^{\obs}$ for all $z \in \mathcal{Z}^1$ almost surely.
\end{assumption}

Assumption \ref{as:first-focal} requires the first-stage focal subpopulation to remain unchanged over $\mathcal{Z}^1$.
For example, this assumption is satisfied by the selection method described in Section \ref{subsec:first-focal}.
Since any $z \in \mathcal{Z}^1$ keeps the treatment assignments of focal units fixed, the first-stage imputable set $\mathcal{Z}^1$ and conditional randomization distribution $\pi_1$ would also remain unchanged even if $(a, z)$ were observed.

Importantly, the definitions of $\mathrm{pval}_1(a, z)$ and $\phi_1(a, z)$ do not require the null hypothesis of no spillover $H_1$ to hold.
However, computing $\mathrm{pval}_1(a, z)$ and $\phi_1(a, z)$ requires imputing $Y_{\mathcal{F}_1^{\obs}}(z)$ from the observed data.
We address this requirement in the selective randomization inference developed below.

\subsection{Selective randomization inference after the first-stage rejection} \label{subsec:reject}

If the first-stage CRT rejects the null hypothesis of no spillover $H_1$ in \eqref{eq:H1}, we move to testing the null hypothesis of zero exposure contrast $H_2$ in \eqref{eq:H2} while assuming the correctness of the exposure mapping $E_i(Z) = (Z_i, A_{c(i)})$ in \eqref{eq:exposure}.
Formally, we impose the next assumption in the first-stage rejection case.

\begin{assumption} \label{as:correct-exposure}
    $E_i(Z) = (Z_i, A_{c(i)})$ is a correct exposure mapping in the sense of \eqref{eq:correct-exposure}.
\end{assumption}

\subsubsection{Imputability of potential outcomes} \label{subsubsec:second-reject-impute}

Under Assumption \ref{as:correct-exposure}, we examine the imputability of potential outcomes under $H_2$.
Take an arbitrary unit $i \in  [N]$.
Denote $E_i^{\obs} \coloneqq E_i(Z^{\obs})$.
If $E_i^{\obs} \in \{ \mathtt{a}, \mathtt{b} \}$, $H_2$ implies that $Y_i(z) = Y_i^{\obs}$ for any $z \in \mathcal{Z}$ such that $E_i(z) \in \{ \mathtt{a}, \mathtt{b} \}$.
Even if $E_i^{\obs} \notin \{ \mathtt{a}, \mathtt{b} \}$, Assumption \ref{as:correct-exposure} ensures that $Y_i(z) = Y_i^{\obs}$ for any $z \in \mathcal{Z}$ such that $E_i(z) = E_i^{\obs}$.
Thus, the largest imputable set of unit $i$ with respect to $H_2$ is given by
\begin{align} \label{eq:mathcal-Z-2}
    \mathcal{Z}_i^2
    \coloneqq 
    \begin{cases*}
       \{ z \in \mathcal{Z}: E_i(z) \in \{ \mathtt{a}, \mathtt{b} \} \} & \text{for $i \in [N]$ such that $E_i^{\obs} \in \{ \mathtt{a}, \mathtt{b} \}$,}\\
       \{ z \in \mathcal{Z}: E_i(z) = E_i^{\obs} \} & \text{for $i \in [N]$ such that $E_i^{\obs} \notin \{ \mathtt{a}, \mathtt{b} \}$.}
    \end{cases*}
\end{align}
This is the largest imputable set for unit $i$ because, for any $z \notin \mathcal{Z}_i^2$, $H_2$ does not uniquely determine $Y_i(z)$ from the observed data.
Accordingly, by Definition \ref{def:partial}, $H_2$ is partially sharp if and only if it is not sharp and $\mathcal{Z}_i^2$ contains at least one assignment other than $Z^{\obs}$ for at least one unit $i \in [N]$.

\begin{example}[Six units in two clusters, continued] \label{example3}
    In the setting of Examples \ref{example1} and \ref{example2}, suppose that Assumption \ref{as:correct-exposure} holds.
    We set $\mathtt{a} = (1, 2/3)$ and $\mathtt{b} = (0, 2/3)$ to test for direct effects under the saturation label $s = 2/3$. 
    Table \ref{tab:example2} shows the imputable potential outcomes with respect to the null hypothesis of zero exposure contrast $H_2$.
    For example, unit 4 has the exposure $E_4^{\obs} = (1, 2/3) \in \{ \mathtt{a}, \mathtt{b} \}$, and the largest imputable set for unit 4 is $\mathcal{Z}_4^2 = \{ z \in \mathcal{Z}: E_4(z) \in \{ \mathtt{a}, \mathtt{b} \} \} = \{ z_{(1)}, \dots, z_{(9)} \}$.
    As another example, unit 3 has the exposure $E_3^{\obs} = (0, 1/3) \notin \{ \mathtt{a}, \mathtt{b} \}$, and the largest imputable set for unit 3 is $\mathcal{Z}_3^2  = \{ z \in \mathcal{Z}: E_3(z) = E_3^{\obs} \} = \{ z_{(1)}, \dots, z_{(6)} \}$.
\end{example}

\begin{table}[!]
    \centering
    \caption{Treatment assignments $z \in \mathcal{Z}$ and imputable potential outcomes $Y_i(z)$ under the null hypothesis of zero exposure contrast $H_2$ with $\mathtt{a} = (1, 2/3)$ and $\mathtt{b} = (0, 2/3)$ when Assumption \ref{as:correct-exposure} is imposed.
    Red question marks indicate potential outcomes that are not imputable. 
    The columns labeled ``\textbf{Focal 2}'' indicate the second-stage focal units $\mathcal{F}_2^{\mathrm{obs}} = \{ 4, 5, 6 \}$ used in Example \ref{example4}, and shaded rows indicate treatment assignments in the imputable set $\mathcal{Z}^2 = \{ Z^{\obs}, z_{(2)}, z_{(4)}, z_{(5)} \}$.}
    \label{tab:example2}
    \begin{tabular}{r|cccccc}
         & \multicolumn{6}{c}{The potential outcome $Y_i(z)$} \\
        \diagbox[width=15em]{Assignment $z$}{Unit $i$} & $i = 1$ & $i = 2$ & $i = 3$ & $\bm{i = 4}$ & $\bm{i = 5}$ & $\bm{i = 6}$ \\ 
        & & & {\scriptsize Focal 1} & {\scriptsize Focal 1} & & \\
        & & & & {\scriptsize \textbf{Focal 2}} & {\scriptsize \textbf{Focal 2}} & {\scriptsize \textbf{Focal 2}} \\
        \hline 
        \multicolumn{1}{l|}{Given $A_1 = 1/3$ and $A_2 = 2/3$:} \\
        \rowcolor{gray!30} $Z^{\obs} = z_{(1)} = (1, 0, 0, 1, 1, 0)$ & 11 & 12 & 13 & 14 & 15 & 16 \\
        \rowcolor{gray!30} $z_{(2)} = (1, 0, 0, 1, 0, 1)$ & 11 & 12 & 13 & 14 & 15 & 16 \\
        $z_{(3)} = (1, 0, 0, 0, 1, 1)$ & 11 & 12 & 13 & 14 & 15 & 16 \\
        \rowcolor{gray!30} $z_{(4)} = (0, 1, 0, 1, 1, 0)$ & \textcolor{red}{?} & \textcolor{red}{?} & 13 & 14 & 15 & 16 \\
        \rowcolor{gray!30} $z_{(5)} = (0, 1, 0, 1, 0, 1)$ & \textcolor{red}{?} & \textcolor{red}{?} & 13 & 14 & 15 & 16 \\
        $z_{(6)} = (0, 1, 0, 0, 1, 1)$ & \textcolor{red}{?} & \textcolor{red}{?} & 13 & 14 & 15 & 16 \\
        $z_{(7)} = (0, 0, 1, 1, 1, 0)$ & \textcolor{red}{?} & 12 & \textcolor{red}{?} & 14 & 15 & 16 \\
        $z_{(8)} = (0, 0, 1, 1, 0, 1)$ & \textcolor{red}{?} & 12 & \textcolor{red}{?} & 14 & 15 & 16 \\
        $z_{(9)} = (0, 0, 1, 0, 1, 1)$ & \textcolor{red}{?} & 12 & \textcolor{red}{?} & 14 & 15 & 16 \\
        \\
        \multicolumn{1}{l|}{Given $A_1 = 2/3$ and $A_2 = 1/3$:} \\
        $z_{(10)} = (1, 1, 0, 1, 0, 0)$ & \textcolor{red}{?} & \textcolor{red}{?} & \textcolor{red}{?} & \textcolor{red}{?} & \textcolor{red}{?} & \textcolor{red}{?} \\
        $z_{(11)} = (1, 1, 0, 0, 1, 0)$ & \textcolor{red}{?} & \textcolor{red}{?} & \textcolor{red}{?} & \textcolor{red}{?} & \textcolor{red}{?} & \textcolor{red}{?} \\
        $z_{(12)} = (1, 1, 0, 0, 0, 1)$ & \textcolor{red}{?} & \textcolor{red}{?} & \textcolor{red}{?} & \textcolor{red}{?} & \textcolor{red}{?} & \textcolor{red}{?} \\
        $z_{(13)} = (1, 0, 1, 1, 0, 0)$ & \textcolor{red}{?} & \textcolor{red}{?} & \textcolor{red}{?} & \textcolor{red}{?} & \textcolor{red}{?} & \textcolor{red}{?} \\
        $z_{(14)} = (1, 0, 1, 0, 1, 0)$ & \textcolor{red}{?} & \textcolor{red}{?} & \textcolor{red}{?} & \textcolor{red}{?} & \textcolor{red}{?} & \textcolor{red}{?} \\
        $z_{(15)} = (1, 0, 1, 0, 0, 1)$ & \textcolor{red}{?} & \textcolor{red}{?} & \textcolor{red}{?} & \textcolor{red}{?} & \textcolor{red}{?} & \textcolor{red}{?} \\
        $z_{(16)} = (0, 1, 1, 1, 0, 0)$ & \textcolor{red}{?} & \textcolor{red}{?} & \textcolor{red}{?} & \textcolor{red}{?} & \textcolor{red}{?} & \textcolor{red}{?} \\
        $z_{(17)} = (0, 1, 1, 0, 1, 0)$ & \textcolor{red}{?} & \textcolor{red}{?} & \textcolor{red}{?} & \textcolor{red}{?} & \textcolor{red}{?} & \textcolor{red}{?} \\
        $z_{(18)} = (0, 1, 1, 0, 0, 1)$ & \textcolor{red}{?} & \textcolor{red}{?} & \textcolor{red}{?} & \textcolor{red}{?} & \textcolor{red}{?} & \textcolor{red}{?} \\
    \end{tabular}
\end{table}

\subsubsection{Selective randomization inference} 

To test the null hypothesis of zero exposure contrast $H_2$, we begin by constructing a focal subpopulation.
Specifically, let $\mathcal{F}_2: \mathcal{Z} \to 2^{[N]}$ be a pre-specified set-valued function such that
\begin{align} \label{eq:second-focal}
    \mathcal{F}_2(z) \subseteq \left\{ i \in [N]: E_i(z) \in \{ \mathtt{a}, \mathtt{b} \} 
    \right\}
    \qquad \text{for all $z \in \mathcal{Z}$.}
\end{align}
The restriction \eqref{eq:second-focal} focuses the second-stage test on comparing the two exposure values $\mathtt{a}$ and $\mathtt{b}$.
As in the first-stage CRT, the selection function $\mathcal{F}_2$ may depend on auxiliary randomness independent of the treatment assignment, and we condition on its realized value and suppress this dependence in the notation.
For each $z \in \mathcal{Z}$, let $F_{2i}(z) \coloneqq \bm{1}\{ i \in \mathcal{F}_2(z) \}$ and $F_2(z) \coloneqq ( F_{2i}(z) : i \in [N] )$.
For the observed assignment, define $\mathcal{F}_2^{\obs} \coloneqq \mathcal{F}_2(Z^{\obs})$, $F_{2i}^{\obs} \coloneqq F_{2i}(Z^{\obs})$, and $F_2^{\obs} \coloneqq F_2(Z^{\obs})$.
Importantly, $\mathcal{F}_2^{\obs}$ need not coincide with $\mathcal{F}_1^{\obs}$, although the two may overlap.

To define the largest imputable set with respect to $H_2$ for $\mathcal{F}_2^{\obs}$, recall the unit-level largest imputable set $\mathcal{Z}_i^2$ defined in \eqref{eq:mathcal-Z-2}.
Since all second-stage focal units $i \in \mathcal{F}_2^{\obs}$ satisfy $E_i^{\obs} \in \{ \mathtt{a}, \mathtt{b} \}$ by \eqref{eq:second-focal}, the largest imputable set for $\mathcal{F}_2^{\obs}$ is given by
\begin{align} \label{eq:largest2}
    \bigcap_{i \in \mathcal{F}_2^{\obs}} \mathcal{Z}_i^2
    = \left\{ z \in \mathcal{Z}: E_i(z) \in \{ \mathtt{a}, \mathtt{b} \} \text{ for all $i \in \mathcal{F}_2^{\obs}$} \right\}.
\end{align}
By construction, for any assignment $z$ in this set, the potential outcome vector $Y_{\mathcal{F}_2^{\obs}}(z)$ is imputable under $H_2$;
that is, $Y_{\mathcal{F}_2^{\obs}}(z) = Y_{\mathcal{F}_2^{\obs}}^{\obs}$ under $H_2$.

Selective randomization inference requires imputing both the first-stage rejection indicator $\phi_1(a, z)$ and the second-stage test statistic, say $T_2(a, z, Y_{\mathcal{F}_2^{\obs}}(z))$, at counterfactual assignments $(a, z)$ beyond the observed assignments $(A^{\obs}, Z^{\obs})$.
These imputations require the imputability of both $Y_{\mathcal{F}_1^{\obs}}(z)$ and $Y_{\mathcal{F}_2^{\obs}}(z)$ under the null hypothesis of zero exposure contrast $H_2$, without assuming the null hypothesis of no spillover $H_1$.
Restricting assignments to the largest imputable set in \eqref{eq:largest2} does not necessarily ensure this, because $\mathcal{F}_1^{\obs}$ may contain units outside $\mathcal{F}_2^{\obs}$.
In addition, we need to restrict assignments to the first-stage largest imputable set $\mathcal{Z}^1$ so that the first-stage rejection indicator $\phi_1(a, z)$ is obtained with the same focal subpopulation $\mathcal{F}_1^{\obs}$ and conditional randomization distribution $\pi_1$ as the observed first-stage rejection indicator $\phi_1^{\obs}$.

Combining these requirements, we define the following imputable set for selective randomization inference to test $H_2$:
\begin{align} \label{eq:mathcal-Z-F1F2}
    \mathcal{Z}_{\mathcal{F}_1^{\obs} \mathcal{F}_2^{\obs}}^2
    \coloneqq \mathcal{Z}_{\mathcal{F}_1^{\obs}}^1 \cap \left( \bigcap_{i \in \mathcal{F}_1^{\obs} \cup \mathcal{F}_2^{\obs}} \mathcal{Z}_i^2 \right).
\end{align}
For notational simplicity, we often write $\mathcal{Z}^2 \coloneqq \mathcal{Z}_{\mathcal{F}_1^{\obs} \mathcal{F}_2^{\obs}}^2$ when no confusion arises.
Equivalently,
\begin{align*}
    \mathcal{Z}^2
    = \{z \in \mathcal{Z}^1:
    & \; E_i(z) \in \{ \mathtt{a}, \mathtt{b} \} \text{ for all $i \in \mathcal{F}_1^{\obs} \cup \mathcal{F}_2^{\obs}$ such that $E_i^{\obs} \in \{ \mathtt{a}, \mathtt{b} \}$,} \\
    & \; E_i(z) = E_i^{\obs} \text{ for all $i \in \mathcal{F}_1^{\obs}$ such that $E_i^{\obs} \notin \{ \mathtt{a}, \mathtt{b} \}$} \}.
\end{align*}
Note that the second restriction applies only to first-stage focal units because all second-stage focal units satisfy $E_i^{\obs} \in \{ \mathtt{a}, \mathtt{b} \}$.
For all $i \in \mathcal{F}_1^{\obs} \cup \mathcal{F}_2^{\obs}$ and $z \in \mathcal{Z}^2$, we have $Y_i(z) = Y_i^{\obs}$ under $H_2$ and Assumption \ref{as:correct-exposure}.
Thus, for each $z \in \mathcal{Z}^2$ and compatible saturation assignment $a$, both the first-stage rejection indicator $\phi_1(a, z)$ and the second-stage test statistic $T_2(a, z, Y_{\mathcal{F}_2^{\obs}}(z))$ are imputable from the observed data.

\begin{example} \label{example4}
    We continue the setting of six units in two clusters with $\mathtt{a} = (1, 2/3)$ and $\mathtt{b} = (0, 2/3)$ in Examples \ref{example1}, \ref{example2}, and \ref{example3}.
    We have $\mathcal{F}_1^{\obs} = \{ 3, 4 \}$ and $\mathcal{Z}^1 = \{ z_{(1)}, z_{(2)}, z_{(4)}, z_{(5)}, z_{(10)} \}$ as described in Example \ref{example2}.
    Only units $i \in \{ 4, 5, 6 \}$ satisfy $E_i(Z^{\obs}) \in \{ \mathtt{a}, \mathtt{b} \}$.
    If we set $\mathcal{F}_2^{\obs} = \{ 4, 5, 6 \}$, then $\mathcal{F}_1^{\obs} \cup \mathcal{F}_2^{\obs} = \{ 3, 4, 5, 6 \}$, and we have 
    \begin{align*}
        \bigcap_{i \in \mathcal{F}_2^{\obs}} \mathcal{Z}_i^2
        & = \{ z_{(1)}, z_{(2)}, z_{(3)}, z_{(4)}, z_{(5)}, z_{(6)}, z_{(7)}, z_{(8)}, z_{(9)} \}, \\
        \bigcap_{i \in \mathcal{F}_1^{\obs} \cup \mathcal{F}_2^{\obs}} \mathcal{Z}_i^2
        & = \{ z_{(1)}, z_{(2)}, z_{(3)}, z_{(4)}, z_{(5)}, z_{(6)} \}.
    \end{align*}
    Thus, the imputable set $\mathcal{Z}^2$ defined in \eqref{eq:mathcal-Z-F1F2} is given by
    \begin{align*}
        \mathcal{Z}^2
        = \mathcal{Z}^1 \cap \bigcap_{i \in \mathcal{F}_1^{\obs} \cup \mathcal{F}_2^{\obs}} \mathcal{Z}_i^2
        = \{ z_{(1)}, z_{(2)}, z_{(4)}, z_{(5)} \},
    \end{align*}
    which corresponds to the shaded rows in Table \ref{tab:example2}.
\end{example}

Given the first-stage rejection $\phi_1^{\obs} = 1$, the focal subpopulations $\mathcal{F}_1^{\obs}$ and $\mathcal{F}_2^{\obs}$, and the imputable set $\mathcal{Z}^2$, the SRT for $H_2$ proceeds as follows.
Let $T_2(a, z, Y_{\mathcal{F}_2^{\obs}}(z))$ be a generic test statistic computed with the treatment assignment $z \in \mathcal{Z}^2$, the corresponding saturation assignment $a$, and the corresponding potential outcome vector $Y_{\mathcal{F}_2^{\obs}}(z)$.
For example, we may consider the absolute difference in means between the focal units exposed to $\mathtt{a}$ and those exposed to $\mathtt{b}$:
\begin{align} \label{eq:second-rejection-statistic}
    T_2(a, z, Y_{\mathcal{F}_2^{\obs}}(z))
    = \left| \frac{ \sum_{i \in \mathcal{F}_2^{\obs}} \bm{1}\{ E_i(z) = \mathtt{a} \} Y_i(z) }{ \sum_{i \in \mathcal{F}_2^{\obs}} \bm{1}\{ E_i(z) = \mathtt{a} \} } - \frac{ \sum_{i \in \mathcal{F}_2^{\obs}} \bm{1}\{ E_i(z) = \mathtt{b} \} Y_i(z) }{ \sum_{i \in \mathcal{F}_2^{\obs}} \bm{1}\{ E_i(z) = \mathtt{b} \} } \right|.
\end{align}
As discussed above, under Assumption \ref{as:correct-exposure}, we can impute the potential outcome vectors $Y_{\mathcal{F}_1^{\obs}}(z) = Y_{\mathcal{F}_1^{\obs}}^{\obs}$ and $Y_{\mathcal{F}_2^{\obs}}(z) = Y_{\mathcal{F}_2^{\obs}}^{\obs}$ for all $z \in \mathcal{Z}^2$ under $H_2$.
Then, the selective $p$-value for testing $H_2$ after the first-stage rejection $\phi_1^{\obs} = 1$ is defined by
\begin{align} \label{eq:pval-second}
\begin{split}
    \mathrm{pval}_2(A^{\obs}, Z^{\obs}) 
    \coloneqq \Pr \Big[ & T_2 \big( A, Z, Y_{\mathcal{F}_2^{\obs}}^{\obs} \big) \ge T_2 \big( A^{\obs}, Z^{\obs}, Y_{\mathcal{F}_2^{\obs}}^{\obs} \big) \; \Big| \; \\
    & \phi_1(A, Z) = 1, F_1(Z) = F_1^{\obs}, F_2(Z) = F_2^{\obs}, Z \in \mathcal{Z}^2 \Big],
\end{split}
\end{align}
where the probability is taken with respect to the conditional distribution of $(A, Z)$ induced by the randomized saturation design, given $\phi_1(A, Z) = 1$, $F_1(Z) = F_1^{\obs}$, $F_2(Z) = F_2^{\obs}$, and $Z \in \mathcal{Z}^2$, with the observed data and the conditioning quantities held fixed.
The SRT given $\phi_1^{\obs} = 1$ rejects $H_2$ at level $\alpha_2 \in [0, 1]$ if $\mathrm{pval}_2(A^{\obs}, Z^{\obs}) \le \alpha_2$.
Algorithm \ref{alg:second-stage-rejection} in Appendix \ref{sec:algoirthm} summarizes this testing procedure.

As the first main result of this paper, the following theorem establishes that the SRT for $H_2$ after the first-stage rejection controls the selective Type I error rate in finite samples, without assuming $H_1$.

\begin{theorem} \label{thm:after-rejection}
    Let Assumptions \ref{as:first-focal} and \ref{as:correct-exposure} hold.
    Suppose that the following three conditions hold in addition to conditions (i)--(iii) in Proposition \ref{prop:stage1}.
    (i) Let $\mathcal{F}_2: \mathcal{Z} \to 2^{[N]}$ be a pre-specified second-stage focal-subpopulation selection function satisfying \eqref{eq:second-focal}, with the observed focal subpopulation $\mathcal{F}_2^{\obs} = \mathcal{F}_2(Z^{\obs})$ and the corresponding indicator vector $F_2^{\obs}$.
    (ii) Let $\mathcal{Z}^2 = \mathcal{Z}_{\mathcal{F}_1^{\obs} \mathcal{F}_2^{\obs}}^2$ be the imputable set as defined in \eqref{eq:mathcal-Z-F1F2}.
    (iii) Let $T_2(a, z, Y_{\mathcal{F}_2^{\obs}}(z))$ be a pre-specified test statistic that is well defined for any treatment assignment $z \in \mathcal{Z}^2$, the corresponding saturation assignment $a$, and the corresponding potential outcome vector $Y_{\mathcal{F}_2^{\obs}}(z)$.
    Then, under $H_2$, for all $\alpha_2 \in [0, 1]$,
    \begin{align*}
        \Pr \left[ \mathrm{pval}_2(A^{\obs}, Z^{\obs}) \le \alpha_2 \;\middle|\; \phi_1^{\obs} = 1, F_1^{\obs}, F_2^{\obs}, \mathcal{Z}^2 \right] \le \alpha_2
        \qquad \text{almost surely},
    \end{align*}
    where the probability is taken with respect to the conditional distribution of $(A^{\obs}, Z^{\obs})$ given $\phi_1^{\obs} = 1$, $F_1^{\obs}$, $F_2^{\obs}$, and $\mathcal{Z}^2$.
\end{theorem}

The selective $p$-value in \eqref{eq:pval-second} can be approximated by sampling counterfactual assignments $(a, z)$ from the following selective randomization distribution:
\begin{align*}
    \pi_2(a, z)
    \coloneqq \Pr \Big(A = a, Z = z \; \Big| \; \phi_1(A, Z) = 1, F_1(Z) = F_1^{\obs}, F_2(Z) = F_2^{\obs}, Z \in \mathcal{Z}^2 \Big).
\end{align*}
While sampling from this distribution can be computationally demanding in general, we can adopt an easy-to-implement Monte Carlo procedure if there are only two saturation labels $\mathcal{S} = \mathcal{S}_{\mathrm{int}} = \{ s_1, s_2 \}$ and we set $\mathtt{a} = (1, s)$ and $\mathtt{b} = (0, s)$ for $s \in \mathcal{S}$.
See Appendix \ref{sec:computation} for details.

\begin{remark} \label{remark:CI1}
    If we assume the homogeneity of the exposure effect $y_i(\mathtt{a}) - y_i(\mathtt{b})$ over $i \in [N]$, we can construct a selective confidence set for the average exposure effect $N^{-1} \sum_{i=1}^N [ y_i(\mathtt{a}) - y_i(\mathtt{b}) ]$ by test inversion.
    Specifically, for a given constant $c \in \mathbb{R}$, we replace $H_2$ with
    \begin{align*}
        H_2(c): \qquad
        y_i(\mathtt{a}) - y_i(\mathtt{b}) = c \quad \text{for all } i \in [N].
    \end{align*}
    Under Assumption \ref{as:correct-exposure}, for any $i \in \mathcal{F}_1^{\obs} \cup \mathcal{F}_2^{\obs}$ and $z \in \mathcal{Z}^2$, $H_2(c)$ implies that
    \begin{align*}
        Y_i(z) = Y_i^{\obs} + c \left[ \bm{1}\{ E_i(z) = \mathtt{a} \} - \bm{1}\{ E_i^{\obs} = \mathtt{a} \} \right].
    \end{align*}
    Using these imputed outcomes, we can evaluate both the first-stage rejection indicator and the second-stage test statistic and perform an SRT at level $\alpha_2$.
    Under the homogeneity of $y_i(\mathtt{a}) - y_i(\mathtt{b})$, the set of values of $c$ for which this test does not reject $H_2(c)$ forms a selective confidence set for the average exposure effect.
    By the same argument as in Theorem \ref{thm:after-rejection}, its selective coverage probability is at least $1 - \alpha_2$, given $\phi_1^{\obs} = 1$, $F_1^{\obs}$, $F_2^{\obs}$, and $\mathcal{Z}^2$.
\end{remark}

\subsubsection{Choice of focal subpopulation} \label{subsubsec:second-reject-focal}

We can construct the second-stage focal subpopulation selection function $\mathcal{F}_2$ using a biclique of the null exposure graph by building on \citet{puelz2022graph}.
In the following, we focus on the construction of $\mathcal{F}_2^{\obs}$, as the function $\mathcal{F}_2$ is defined by applying the same construction to each hypothetical observed assignment $z \in \mathcal{Z}$, replacing $Z^{\obs}$ by $z$ throughout and recomputing all assignment-dependent quantities.
Then, the selective $p$-value is computed as in \eqref{eq:pval-second}.

To proceed, given the first-stage focal subpopulation $\mathcal{F}_1^{\obs}$, let
\begin{align} \label{eq:second-biclique-base}
    \mathcal{Z}_{\mathcal{F}_1^{\obs}}^2
    & \coloneqq \mathcal{Z}_{\mathcal{F}_1^{\obs}}^1 \cap \left( \bigcap_{i \in \mathcal{F}_1^{\obs}} \mathcal{Z}_i^2 \right).
\end{align}
Equivalently, 
\begin{align*}
    \mathcal{Z}_{\mathcal{F}_1^{\obs}}^2
    = \{z \in \mathcal{Z}_{\mathcal{F}_1^{\obs}}^1:
    & \; E_i(z) \in \{ \mathtt{a}, \mathtt{b} \} \text{ for all $i \in \mathcal{F}_1^{\obs}$ such that $E_i^{\obs} \in \{ \mathtt{a}, \mathtt{b} \}$,} \\
    & \; E_i(z) = E_i^{\obs} \text{ for all $i \in \mathcal{F}_1^{\obs}$ such that $E_i^{\obs} \notin \{ \mathtt{a}, \mathtt{b} \}$} \}.
\end{align*}
This set contains the observed assignment $Z^{\obs}$, and the potential outcome vector $Y_{\mathcal{F}_1^{\obs}}(z)$ for any $z$ in this set is imputable under $H_2$ and Assumption \ref{as:correct-exposure}.

The null exposure graph of $H_2$ after the first-stage rejection is defined as a bipartite graph $\bm{G}$ with units $i \in [N]$ on one side and assignments $z \in \mathcal{Z}_{\mathcal{F}_1^{\obs}}^2$ on the other, and an edge connects $i$ and $z$ if and only if $E_i(z) \in \{ \mathtt{a}, \mathtt{b} \}$.
Formally, the null exposure graph is defined as
\begin{align*}
    \bm{G} 
    \coloneqq \left( [N], \mathcal{Z}_{\mathcal{F}_1^{\obs}}^2, \bm{E} \right),
\end{align*}
where $\bm{E}$ is the set of edges defined as
\begin{align*}
    \bm{E}
    \coloneqq \left\{ (i, z) \in [N] \times \mathcal{Z}_{\mathcal{F}_1^{\obs}}^2 : E_i(z) \in \{ \mathtt{a}, \mathtt{b} \} \right\}.
\end{align*}

A biclique of the null exposure graph $\bm{G}$ is a complete bipartite subgraph in which every selected unit $i \in [N]$ is connected to every selected assignment $z \in \mathcal{Z}_{\mathcal{F}_1^{\obs}}^2$.
Specifically, letting $\mathcal{U} \subseteq [N]$ be a set of units, define
\begin{align} \label{eq:second-biclique-neighborhood}
    \mathcal{Z}_{\mathcal{F}_1^{\obs} \mathcal{U}}^2
    \coloneqq \left\{ z \in \mathcal{Z}_{\mathcal{F}_1^{\obs}}^2: E_i(z) \in \{ \mathtt{a}, \mathtt{b} \} \text{ for all $i \in \mathcal{U}$} \right\}.
\end{align}
If both $\mathcal{U}$ and $\mathcal{Z}_{\mathcal{F}_1^{\obs} \mathcal{U}}^2$ are nonempty, then the pair $\bm{B} \coloneqq ( \mathcal{U}, \mathcal{Z}_{\mathcal{F}_1^{\obs} \mathcal{U}}^2 )$ forms a biclique of $\bm{G}$. 
Moreover, if $Z^{\obs} \in \mathcal{Z}_{\mathcal{F}_1^{\obs} \mathcal{U}}^2$, then $E_i^{\obs} \in \{ \mathtt{a}, \mathtt{b} \}$ for all $i \in \mathcal{U}$.
Thus, for any nonempty $\mathcal{U}$ such that $Z^{\obs} \in \mathcal{Z}_{\mathcal{F}_1^{\obs} \mathcal{U}}^2$, we can set $\mathcal{F}_2^{\obs} = \mathcal{U}$ and $\mathcal{Z}_{\mathcal{F}_1^{\obs} \mathcal{F}_2^{\obs}}^2 = \mathcal{Z}_{\mathcal{F}_1^{\obs} \mathcal{U}}^2$.
If there are multiple bicliques containing $Z^{\obs}$, we can choose a biclique that maximizes the number of edges $|\mathcal{U}| \cdot |\mathcal{Z}_{\mathcal{F}_1^{\obs} \mathcal{U}}^2|$ over the candidates containing $Z^{\obs}$.

\begin{remark} \label{remark:biclique}
    When $\mathtt{a} = (1, s)$ and $\mathtt{b} = (0, s)$ for some $s \in \mathcal{S}_{\mathrm{int}}$, first-stage focal units cannot switch between $\mathtt{a}$ and $\mathtt{b}$ under conditional randomization because their own treatment
    assignments are held fixed.
    We may therefore restrict the candidate second-stage focal units to $[N] \setminus \mathcal{F}_1^{\obs}$.
    Specifically, we may replace $\bm{G}$ and $\bm{E}$ with the restricted graph
    \begin{align*}
        \bm{G}^{-}
        \coloneqq \left( [N] \setminus \mathcal{F}_1^{\obs}, \mathcal{Z}_{\mathcal{F}_1^{\obs}}^2, \bm{E}^{-} \right),
        \qquad 
        \bm{E}^{-}
        \coloneqq \left\{ (i, z) \in ([N] \setminus \mathcal{F}_1^{\obs}) \times \mathcal{Z}_{\mathcal{F}_1^{\obs}}^2: E_i(z) \in \{\mathtt{a},\mathtt{b}\} \right\}.
    \end{align*}
    We can then apply the biclique method described above to this restricted graph.
\end{remark}

\subsection{Selective randomization inference after the first-stage non-rejection} \label{subsec:accept}

If the first-stage CRT does not reject the null hypothesis of no spillover $H_1$ in \eqref{eq:H1}, we move to testing the null hypothesis of no direct effects $\tilde{H}_1$ in \eqref{eq:tildeH1} under the assumption of no treatment spillover.
Formally, we impose the next assumption in the first-stage non-rejection case.

\begin{assumption} \label{as:no-spillover}
    There is no treatment spillover in the sense of \eqref{eq:H1}.
\end{assumption}

\subsubsection{Imputability of potential outcomes}

Under Assumption \ref{as:no-spillover}, $\tilde{H}_1$ imputes all potential outcomes, that is, $Y_i(z) = Y_i^{\obs}$ for all $i \in [N]$ and $z \in \mathcal{Z}$.
Thus, the largest imputable set for each unit with respect to $\tilde{H}_1$ is the entire support $\mathcal{Z}$.
Although all potential outcomes are imputable over $\mathcal{Z}$, our testing procedure conditions on the first-stage focal subpopulation and its observed treatment assignments, restricting randomization to $\mathcal{Z}^1$.
Under Assumption \ref{as:first-focal}, this restriction keeps the first-stage focal subpopulation $\mathcal{F}_1^{\obs}$ and conditional randomization distribution $\pi_1$ unchanged, so that $\phi_1(a, z)$ reproduces the counterfactual first-stage test result under any $z \in \mathcal{Z}^1$ and the corresponding $a$.

Given this, the SRT for $\tilde{H}_1$ sets the imputable set and the focal-subpopulation selection function as follows.
Given the first-stage focal subpopulation $\mathcal{F}_1^{\obs}$, we set the imputable set equal to the first-stage imputable set:
\begin{align} \label{eq:tilde-mathcal-Z-F1}
    \tilde{\mathcal{Z}}_{\mathcal{F}_1^{\obs}}^1
    \coloneqq \mathcal{Z}_{\mathcal{F}_1^{\obs}}^1
    = \{ z \in \mathcal{Z}: z_i = Z_i^{\obs} \text{ for all $i \in \mathcal{F}_1^{\obs}$} \}.
\end{align}
For notational simplicity, we often write $\tilde{\mathcal{Z}}^1 \coloneqq \tilde{\mathcal{Z}}_{\mathcal{F}_1^{\obs}}^1$ when no confusion arises.
Within this set, the treatment assignments of the first-stage focal units remain fixed.
To use the remaining variation in treatment assignments, we take the complementary subpopulation as the focal subpopulation for testing $\tilde{H}_1$.
Specifically, we consider the following focal-subpopulation selection function $\tilde{\mathcal{F}}_1: \mathcal{Z} \to 2^{[N]}$ for testing $\tilde{H}_1$:
\begin{align} \label{eq:second-accept-focal}
    \tilde{\mathcal{F}}_1(z) = [N] \setminus \mathcal{F}_1(z)
    \qquad 
    \text{for all $z \in \mathcal{Z}$}.
\end{align}
For each $z \in \mathcal{Z}$, define $\tilde{F}_{1i}(z) \coloneqq \bm{1}\{ i \in \tilde{\mathcal{F}}_1(z) \}$ and $\tilde{F}_1(z) \coloneqq (\tilde{F}_{1i}(z): i \in [N])$.
Note that $\tilde{F}_{1i}(z) = 1 - F_{1i}(z)$.
For the observed assignment, let $\tilde{\mathcal{F}}_1^{\obs} \coloneqq \tilde{\mathcal{F}}_1(Z^{\obs})$, $\tilde{F}_{1i}^{\obs} \coloneqq \bm{1}\{ i \in \tilde{\mathcal{F}}_1^{\obs} \}$, and $\tilde{F}_1^{\obs} \coloneqq (\tilde{F}_{1i}^{\obs}: i \in [N])$.
By construction, both the potential outcome vector $Y_{\tilde{\mathcal{F}}_1^{\obs}}(z) \coloneqq ( Y_i(z): i \in \tilde{\mathcal{F}}_1^{\obs} )$ and the first-stage rejection indicator $\phi_1(a, z)$ are imputable under $\tilde{H}_1$ for any treatment assignment $z \in \tilde{\mathcal{Z}}^1$ and the corresponding saturation assignment $a$.

\begin{example} \label{example5}
    We continue Examples \ref{example1}, \ref{example2}, \ref{example3}, and \ref{example4}.
    Under Assumption \ref{as:no-spillover}, Table \ref{tab:example3} shows that $\tilde{H}_1$ imputes all potential outcomes.
    As in Example \ref{example2}, the first-stage CRT uses $\mathcal{F}_1^{\obs} = \{ 3, 4 \}$ and $\mathcal{Z}^1 = \{ Z^{\obs}, z_{(2)}, z_{(4)}, z_{(5)}, z_{(10)} \}$.
    In the SRT for $\tilde{H}_1$, we set the imputable set and the focal subpopulation as $\tilde{\mathcal{Z}}^1 = \mathcal{Z}^1$ and $\tilde{\mathcal{F}}_1^{\obs} = [N] \setminus \mathcal{F}_1^{\obs} = \{ 1, 2, 5, 6 \}$, which correspond to the shaded rows and the columns labeled ``$\tilde{\text{\textbf{Focal}}}$ \textbf{1}'', respectively, in Table \ref{tab:example3}. 
    Clearly, we can impute both $Y_{\tilde{\mathcal{F}}_1^{\obs}}(z)$ and $\phi_1(a, z)$ under $\tilde{H}_1$ for any $z \in \tilde{\mathcal{Z}}^1$ and the corresponding $a$.
\end{example}

\begin{table}[t]
    \centering
    \caption{Treatment assignments $z \in \mathcal{Z}$ and imputable potential outcomes $Y_i(z)$ under the null hypothesis of no direct effects $\tilde{H}_1$ when Assumption \ref{as:no-spillover} is imposed.
    All potential outcomes are imputable.
    The columns labeled ``$\tilde{\text{\textbf{Focal}}}$ \textbf{1}'' indicate the focal units $\tilde{\mathcal{F}}_1^{\mathrm{obs}} = \{ 1, 2, 5, 6 \}$ used in Example \ref{example5}, and the shaded rows indicate treatment assignments in the imputable set $\tilde{\mathcal{Z}}^1 = \{ Z^{\obs}, z_{(2)}, z_{(4)}, z_{(5)}, z_{(10)} \}$.
    }
    \label{tab:example3}
    \begin{tabular}{r|cccccc}
         & \multicolumn{6}{c}{The potential outcome $Y_i(z)$} \\
        \diagbox[width=15em]{Assignment $z$}{Unit $i$} & $\bm{i = 1}$ & $\bm{i = 2}$ & $i = 3$ & $i = 4$ & $\bm{i = 5}$ & $\bm{i = 6}$ \\ 
        & & & {\scriptsize Focal 1} & {\scriptsize Focal 1} & & \\
        & {\scriptsize $\tilde{\text{\textbf{Focal}}}$ \textbf{1}} & {\scriptsize $\tilde{\text{\textbf{Focal}}}$ \textbf{1}} & & & {\scriptsize $\tilde{\text{\textbf{Focal}}}$ \textbf{1}} & {\scriptsize $\tilde{\text{\textbf{Focal}}}$ \textbf{1}} \\
        \hline 
        \multicolumn{1}{l|}{Given $A_1 = 1/3$ and $A_2 = 2/3$:} \\
        \rowcolor{gray!30} $Z^{\obs} = z_{(1)} = (1, 0, 0, 1, 1, 0)$ & 11 & 12 & 13 & 14 & 15 & 16 \\
        \rowcolor{gray!30} $z_{(2)} = (1, 0, 0, 1, 0, 1)$ & 11 & 12 & 13 & 14 & 15 & 16 \\
        $z_{(3)} = (1, 0, 0, 0, 1, 1)$ & 11 & 12 & 13 & 14 & 15 & 16 \\
        \rowcolor{gray!30} $z_{(4)} = (0, 1, 0, 1, 1, 0)$ & 11 & 12 & 13 & 14 & 15 & 16 \\
        \rowcolor{gray!30} $z_{(5)} = (0, 1, 0, 1, 0, 1)$ & 11 & 12 & 13 & 14 & 15 & 16 \\
        $z_{(6)} = (0, 1, 0, 0, 1, 1)$ & 11 & 12 & 13 & 14 & 15 & 16 \\
        $z_{(7)} = (0, 0, 1, 1, 1, 0)$ & 11 & 12 & 13 & 14 & 15 & 16 \\
        $z_{(8)} = (0, 0, 1, 1, 0, 1)$ & 11 & 12 & 13 & 14 & 15 & 16 \\
        $z_{(9)} = (0, 0, 1, 0, 1, 1)$ & 11 & 12 & 13 & 14 & 15 & 16 \\
        \\
        \multicolumn{1}{l|}{Given $A_1 = 2/3$ and $A_2 = 1/3$:} \\
        \rowcolor{gray!30} $z_{(10)} = (1, 1, 0, 1, 0, 0)$ & 11 & 12 & 13 & 14 & 15 & 16 \\
        $z_{(11)} = (1, 1, 0, 0, 1, 0)$ & 11 & 12 & 13 & 14 & 15 & 16 \\
        $z_{(12)} = (1, 1, 0, 0, 0, 1)$ & 11 & 12 & 13 & 14 & 15 & 16 \\
        $z_{(13)} = (1, 0, 1, 1, 0, 0)$ & 11 & 12 & 13 & 14 & 15 & 16 \\
        $z_{(14)} = (1, 0, 1, 0, 1, 0)$ & 11 & 12 & 13 & 14 & 15 & 16 \\
        $z_{(15)} = (1, 0, 1, 0, 0, 1)$ & 11 & 12 & 13 & 14 & 15 & 16 \\
        $z_{(16)} = (0, 1, 1, 1, 0, 0)$ & 11 & 12 & 13 & 14 & 15 & 16 \\
        $z_{(17)} = (0, 1, 1, 0, 1, 0)$ & 11 & 12 & 13 & 14 & 15 & 16 \\
        $z_{(18)} = (0, 1, 1, 0, 0, 1)$ & 11 & 12 & 13 & 14 & 15 & 16 \\
    \end{tabular}
\end{table}

\subsubsection{Selective randomization inference}

Given the first-stage non-rejection $\phi_1^{\obs} = 0$, the focal subpopulations $\mathcal{F}_1^{\obs}$ and $\tilde{\mathcal{F}}_1^{\obs}$, and the imputable set $\tilde{\mathcal{Z}}^1$, the SRT for $\tilde{H}_1$ proceeds as follows.
Let $\tilde{T}_1(a, z, Y_{\tilde{\mathcal{F}}_1^{\obs}}(z))$ be a generic test statistic computed with the treatment assignment $z \in \tilde{\mathcal{Z}}^1$, the corresponding saturation assignment $a$, and the corresponding potential outcome vector $Y_{\tilde{\mathcal{F}}_1^{\obs}}(z)$.
For example, we may consider the absolute difference in means between the treated and control units:
\begin{align} \label{eq:second-nonrejection-statistic}
    \tilde{T}_1(a, z, Y_{\tilde{\mathcal{F}}_1^{\obs}}(z))
    = \left| \frac{ \sum_{i \in \tilde{\mathcal{F}}_1^{\obs}} \bm{1}\{ z_i = 1 \} Y_i(z) }{ \sum_{i \in \tilde{\mathcal{F}}_1^{\obs}} \bm{1}\{ z_i = 1 \} } - \frac{ \sum_{i \in \tilde{\mathcal{F}}_1^{\obs}} \bm{1}\{ z_i = 0 \} Y_i(z) }{ \sum_{i \in \tilde{\mathcal{F}}_1^{\obs}}  \bm{1}\{ z_i = 0 \} } \right|.
\end{align}
As discussed above, under Assumption \ref{as:no-spillover}, $\tilde{H}_1$ imputes $Y_{\tilde{\mathcal{F}}_1^{\obs}}(z)$ and $\phi_1(a, z)$ for all $z \in \tilde{\mathcal{Z}}^1$ and the corresponding $a$.
Then, the selective $p$-value for testing $\tilde{H}_1$ after the first-stage non-rejection $\phi_1^{\obs} = 0$ is defined by
\begin{align} \label{eq:pval-tildeH1}
\begin{split}
    & \tilde{\mathrm{pval}}_1(A^{\obs}, Z^{\obs}) \\
    & \coloneqq \Pr \Big[ \tilde{T}_1 \big( A, Z, Y_{\tilde{\mathcal{F}}_1^{\obs}}^{\obs} \big) \ge \tilde{T}_1 \big( A^{\obs}, Z^{\obs}, Y_{\tilde{\mathcal{F}}_1^{\obs}}^{\obs} \big) \; \Big| \; \phi_1(A, Z) = 0, \tilde{F}_1(Z) = \tilde{F}_1^{\obs}, Z \in \tilde{\mathcal{Z}}^1 \Big],
\end{split}
\end{align}
where the probability is taken with respect to the conditional distribution of $(A, Z)$ induced by the randomized saturation design, given $\phi_1(A, Z) = 0$, $\tilde{F}_1(Z) = \tilde{F}_1^{\obs}$, and $Z \in \tilde{\mathcal{Z}}^1$, with the observed data and the conditioning quantities held fixed.
Here, conditioning on $\tilde{F}_1(Z) = \tilde{F}_1^{\obs}$ is equivalent to conditioning on $F_1(Z) = F_1^{\obs}$.
Thus, no additional conditioning on $F_1(Z)$ is needed.
The SRT given $\phi_1^{\obs} = 0$ rejects $\tilde{H}_1$ at level $\tilde{\alpha}_1 \in [0, 1]$ if $\tilde{\mathrm{pval}}_1(A^{\obs}, Z^{\obs}) \le \tilde{\alpha}_1$.
Algorithm \ref{alg:second-stage-non-rejection} in Appendix \ref{sec:algoirthm} summarizes this testing procedure.

The following theorem presents the second main result of this paper.

\begin{theorem} \label{thm:after-accept}
    Let Assumptions \ref{as:first-focal} and \ref{as:no-spillover} hold.
    Suppose that the following three conditions hold in addition to conditions (i)--(iii) in Proposition \ref{prop:stage1}.
    (i) Let $\tilde{\mathcal{F}}_1: \mathcal{Z} \to 2^{[N]}$ be a pre-specified focal-subpopulation selection function satisfying \eqref{eq:second-accept-focal}, with the observed focal subpopulation $\tilde{\mathcal{F}}_1^{\obs} = \tilde{\mathcal{F}}_1(Z^{\obs})$ and the corresponding indicator vector $\tilde{F}_1^{\obs}$.
    (ii) Let $\tilde{\mathcal{Z}}^1 = \tilde{\mathcal{Z}}_{\mathcal{F}_1^{\obs}}^1$ be the imputable set as defined in \eqref{eq:tilde-mathcal-Z-F1}.
    (iii) Let $\tilde{T}_1(a, z, Y_{\tilde{\mathcal{F}}_1^{\obs}}(z))$ be a pre-specified test statistic that is well defined for any treatment assignment $z \in \tilde{\mathcal{Z}}^1$, the corresponding saturation assignment $a$, and the corresponding potential outcome vector $Y_{\tilde{\mathcal{F}}_1^{\obs}}(z)$.
    Then, under $\tilde{H}_1$, for all $\tilde{\alpha}_1 \in [0, 1]$,
    \begin{align*}
        \Pr \left[ \tilde{\mathrm{pval}}_1(A^{\obs}, Z^{\obs}) \le \tilde{\alpha}_1 \;\middle|\; \phi_1^{\obs} = 0, \tilde{F}_1^{\obs}, \tilde{\mathcal{Z}}^1 \right] \le \tilde{\alpha}_1
        \qquad \text{almost surely},
    \end{align*}
    where the probability is taken with respect to the conditional distribution of $(A^{\obs}, Z^{\obs})$ given $\phi_1^{\obs} = 0$, $\tilde{F}_1^{\obs}$, and $\tilde{\mathcal{Z}}^1$.
\end{theorem}

The selective $p$-value in \eqref{eq:pval-tildeH1} is based on the following selective randomization distribution:
\begin{align*}
    \tilde{\pi}_1(a, z)
    \coloneqq \Pr \Big(A = a, Z = z \; \Big| \; \phi_1(A, Z) = 0, \tilde{F}_1(Z) = \tilde{F}_1^{\obs}, Z \in \tilde{\mathcal{Z}}^1 \Big).
\end{align*}
In Appendix \ref{sec:computation}, we describe how to approximate the selective $p$-value using the first-stage Monte Carlo sample.

\section{Monte Carlo Simulation} \label{sec:simulation}

In this section, we conduct Monte Carlo experiments to evaluate the size and power of the proposed CRT and SRTs.
In Appendix \ref{sec:simulation2}, we also demonstrate that the conventional conditional randomization test for $H_2$ and the canonical FRT can exhibit distortions in their selective Type I error rates when applied after the first-stage test without accounting for the selection event.

\subsection{Simulation design}

There are $C = 40$ clusters, each of which consists of $N_c = 3$ units.
For the first-stage randomization, we consider two saturation labels $\mathcal{S} = \{ 1/3, 2/3 \}$ and assign each $s \in \mathcal{S}$ to $C/2 = 20$ clusters.
For the second-stage randomization, we assign treatment to $N_{1c}(s) = s N_c$ units in each cluster $c$ under the saturation assignment $A_c = s$, that is, $N_{1c}(1/3) = 1$ and $N_{1c}(2/3) = 2$.

We generate the potential outcomes according to
\begin{align*}
    Y_i(Z)
    = Z_i \beta_1 + \bm{1}\{ A_{c(i)} = 2/3 \} \gamma_1 + Z_i \bm{1}\{ A_{c(i)} = 2/3 \} \delta_1 + \varepsilon_i + \zeta_{c(i)},
\end{align*}
where $(\beta_1, \gamma_1, \delta_1)$ are non-negative coefficients, and $\varepsilon_i$ and $\zeta_c$ are the unit-level and cluster-level error terms, respectively, such that $\varepsilon_i \overset{\text{IID}}{\sim} \mathrm{Normal}(0, \sigma_1^2)$ for $\sigma_1^2 > 0$, $\zeta_c \overset{\text{IID}}{\sim} \mathrm{Normal}(0, \sigma_2^2)$ for $\sigma_2^2 > 0$, and $\{ \varepsilon_i \}_{i=1}^N$ and $\{ \zeta_c \}_{c=1}^C$ are mutually independent.
We specify the parameter values separately for each experiment below.

For the null hypothesis of zero exposure contrast $H_2$, we set $\mathtt{a} = (1, 1/3)$ and $\mathtt{b} = (0, 1/3)$.
The relationship between the coefficients $(\beta_1, \gamma_1, \delta_1)$ and the null hypotheses can be summarized as follows:
\begin{enumerate}[(i)]
    \item $H_1$ is true if and only if $\gamma_1 = \delta_1 = 0$.
    \item $H_2$ is true if and only if $\beta_1 = 0$.
    \item Both $H_1$ and $\tilde H_1$ are true if and only if $\beta_1 = \gamma_1 = \delta_1 = 0$.
\end{enumerate}

Throughout the simulation experiment, we use 10,000 Monte Carlo replications for each parameter configuration.
We evaluate the tests using their rejection rates at the nominal significance levels $\alpha_1 = \alpha_2 = \tilde{\alpha}_1 = 0.1$.

\subsection{First-stage conditional randomization test}

Table \ref{tab:simulation1} presents the simulation results of the first-stage CRT for the null hypothesis of no spillover $H_1$.
To construct the focal subpopulation $\mathcal{F}_1^{\obs}$, we employ the method described in Section \ref{subsec:first-focal} with the number of focal units $K_c = 1$ for any cluster $c$.
For the choice of the test statistic, we use the Wald statistic for testing $\gamma_1 = \delta_1 = 0$ in the linear regression model \eqref{eq:first-statistic}.

Panels (a) and (b) report the empirical Type I error rates and power, respectively, at the level $\alpha_1 = 0.1$ for the parameter configurations $\beta_1 = 0$, $\gamma_1 = \delta_1 \in \{ 0, 0.25, 0.5, 1 \}$, and $\sigma_1^2 = \sigma_2^2 = 0.5$.
We fix the value of $\beta_1$ since the rejection rates of the first-stage CRT are unchanged across values of $\beta_1$.
This is because the treatment assignments of focal units are held fixed under conditional randomization and included as a regressor, so varying $\beta_1$ leaves the Wald statistic unchanged.

Table \ref{tab:simulation1} shows the desirable size and power properties of the first-stage CRT.
When $\gamma_1 = \delta_1 = 0$ so that $H_1$ holds, the empirical rejection rates are close to the nominal level of $\alpha_1 = 0.1$, consistent with Proposition \ref{prop:stage1}.
In contrast, for positive $\gamma_1$ and $\delta_1$ implying that $H_1$ is false, the empirical rejection rates steadily increase as $\gamma_1$ and $\delta_1$ get larger.

\begin{table}[!t]
    \caption{Simulation results for the first-stage conditional randomization test at level $\alpha_1 = 0.1$} \label{tab:simulation1}
    \begin{subtable}[t]{\textwidth}
        \caption{Type I error rate}
        \centering
        \begin{tabular}[t]{rrrrrrrrrr}
            \toprule
            $C$ & $\beta_1$ & $\gamma_1$ & $\delta_1$ & $\sigma_1^2$ & $\sigma_2^2$ & Rejection rate & \# Focal & \# Randomizations & \# MC \\
            \midrule
            40 & 0.00 & 0 & 0 & 0.5 & 0.5 & 0.100 & 40 & 2000 & 10000\\
            \bottomrule
        \end{tabular}
    \end{subtable}
    \par \bigskip 
    \begin{subtable}[t]{\textwidth}
        \caption{Power}
        \centering
        \begin{tabular}[t]{rrrrrrrrrr}
            \toprule
            $C$ & $\beta_1$ & $\gamma_1$ & $\delta_1$ & $\sigma_1^2$ & $\sigma_2^2$ & Rejection rate & \# Focal & \# Randomizations & \# MC \\
            \midrule
            40 & 0.00 & 0.25 & 0.25 & 0.5 & 0.5 & 0.223 & 40 & 2000 & 10000\\
            40 & 0.00 & 0.50 & 0.50 & 0.5 & 0.5 & 0.573 & 40 & 2000 & 10000\\
            40 & 0.00 & 1.00 & 1.00 & 0.5 & 0.5 & 0.979 & 40 & 2000 & 10000\\
            \bottomrule
        \end{tabular}
    \end{subtable}
    \par \bigskip 
    \begin{small}
        Note: 
        The columns ``\# Focal,'' ``\# Randomizations,'' and ``\# MC'' report the numbers of focal units, randomization draws, and Monte Carlo replications, respectively.
    \end{small}
\end{table}

\subsection{Second-stage selective randomization inference}

Tables \ref{tab:simulation21} and \ref{tab:simulation20} present the simulation results of the second-stage SRTs for the null hypothesis of zero exposure contrast $H_2$ and the null hypothesis of no direct effects $\tilde{H}_1$ after the first-stage rejection and non-rejection, respectively, at the levels $\alpha_2 = \tilde{\alpha}_1 = 0.1$.
In each table, panel (a) reports empirical selective Type I error rates, while panel (b) reports rejection rates under alternative parameter configurations.
We employ the same parameter configurations as in Table \ref{tab:simulation1} except that $\beta_1 \in \{ 0, 0.25, 0.5, 1 \}$.
For the focal subpopulation, we use $\mathcal{F}_2^{\obs} = \{ i \in [N]: E_i^{\obs} \in \{ \mathtt{a}, \mathtt{b} \} \} \setminus \mathcal{F}_1^{\obs}$ in the first-stage rejection case, which is the one obtained from the biclique method described in Remark \ref{remark:biclique}, and $\tilde{\mathcal{F}}_1^{\obs} = [N] \setminus \mathcal{F}_1^{\obs}$ in \eqref{eq:second-accept-focal} in the first-stage non-rejection case.
For the test statistics, we employ the absolute difference-in-means statistics in \eqref{eq:second-rejection-statistic} and \eqref{eq:second-nonrejection-statistic}.

Tables \ref{tab:simulation21} and \ref{tab:simulation20} show that the SRTs have satisfactory selective size control and power.
In panels (a) of Tables \ref{tab:simulation21} and \ref{tab:simulation20}, the empirical selective Type I error rates are sufficiently close to the nominal levels $\alpha_2 = \tilde{\alpha}_1 = 0.1$, which is consistent with Theorems \ref{thm:after-rejection} and \ref{thm:after-accept}.
In panel (b) of Table \ref{tab:simulation21}, the power against alternatives to $H_2$ increases with $\beta_1$ for each fixed pair
$(\gamma_1, \delta_1)$.
In panel (b) of Table \ref{tab:simulation20}, the same pattern holds for $\tilde{H}_1$ when $\gamma_1 = \delta_1 = 0$.
Note that the remaining rows in panel (b) of Table \ref{tab:simulation20} report rejection rates when the no-spillover assumption in Assumption \ref{as:no-spillover} is violated.

\begin{table}[!t]
    \caption{Simulation results for the second-stage SRT for $H_2$ after the first-stage rejection at level $\alpha_2 = 0.1$} \label{tab:simulation21}
    \begin{subtable}[t]{\textwidth}
        \caption{Type I error rates}
        \centering
        \begin{tabular}[t]{rrrrrrrrrr}
            \toprule
            $C$ & $\beta_1$ & $\gamma_1$ & $\delta_1$ & $\sigma_1^2$ & $\sigma_2^2$ & Rejection rate & \# Focal & \# Randomizations & \# MC \\
            \midrule
            40 & 0 & 0.00 & 0.00 & 0.5 & 0.5 & 0.103 & 40 & 2000 & 1001\\
            40 & 0 & 0.25 & 0.25 & 0.5 & 0.5 & 0.097 & 40 & 2000 & 2226\\
            40 & 0 & 0.50 & 0.50 & 0.5 & 0.5 & 0.096 & 40 & 2000 & 5726\\
            40 & 0 & 1.00 & 1.00 & 0.5 & 0.5 & 0.098 & 40 & 2000 & 9794\\
            \bottomrule
        \end{tabular}
    \end{subtable}
    \par \bigskip 
    \begin{subtable}[t]{\textwidth}
        \caption{Power}
        \centering
        \begin{tabular}[t]{rrrrrrrrrr}
            \toprule
            $C$ & $\beta_1$ & $\gamma_1$ & $\delta_1$ & $\sigma_1^2$ & $\sigma_2^2$ & Rejection rate & \# Focal & \# Randomizations & \# MC \\
            \midrule
            40 & 0.25 & 0.00 & 0.00 & 0.5 & 0.5 & 0.230 & 40 & 2000 & 1001\\
            40 & 0.25 & 0.25 & 0.25 & 0.5 & 0.5 & 0.255 & 40 & 2000 & 2226\\
            40 & 0.25 & 0.50 & 0.50 & 0.5 & 0.5 & 0.251 & 40 & 2000 & 5726\\
            40 & 0.25 & 1.00 & 1.00 & 0.5 & 0.5 & 0.235 & 40 & 2000 & 9794\\
            40 & 0.50 & 0.00 & 0.00 & 0.5 & 0.5 & 0.513 & 40 & 2000 & 1001\\
            40 & 0.50 & 0.25 & 0.25 & 0.5 & 0.5 & 0.575 & 40 & 2000 & 2226\\
            40 & 0.50 & 0.50 & 0.50 & 0.5 & 0.5 & 0.572 & 40 & 2000 & 5726\\
            40 & 0.50 & 1.00 & 1.00 & 0.5 & 0.5 & 0.549 & 40 & 2000 & 9794\\
            40 & 1.00 & 0.00 & 0.00 & 0.5 & 0.5 & 0.944 & 40 & 2000 & 1001\\
            40 & 1.00 & 0.25 & 0.25 & 0.5 & 0.5 & 0.961 & 40 & 2000 & 2226\\
            40 & 1.00 & 0.50 & 0.50 & 0.5 & 0.5 & 0.962 & 40 & 2000 & 5726\\
            40 & 1.00 & 1.00 & 1.00 & 0.5 & 0.5 & 0.959 & 40 & 2000 & 9794\\
            \bottomrule
        \end{tabular}
    \end{subtable}
    \par \bigskip 
    \begin{small}
        Notes: 
        The column ``\# Focal'' reports the number of focal units.
        Rejection rates are computed using only the Monte Carlo replications with the corresponding first-stage selection, and the column ``\# MC'' reports the number of these replications.
        The column ``\# Randomizations'' reports the average number of randomization draws used in the second-stage test across these replications.
    \end{small}
\end{table}

\begin{table}[t]
    \caption{Simulation results for the second-stage SRT for $\tilde{H}_1$ after the first-stage non-rejection at level $\tilde{\alpha}_1 = 0.1$} \label{tab:simulation20}
    \begin{subtable}[t]{\textwidth}
        \caption{Type I error rate}
        \centering
        \begin{tabular}[t]{rrrrrrrrrr}
            \toprule
            $C$ & $\beta_1$ & $\gamma_1$ & $\delta_1$ & $\sigma_1^2$ & $\sigma_2^2$ & Rejection rate & \# Focal & \# Randomizations & \# MC \\
            \midrule
            40 & 0 & 0 & 0 & 0.5 & 0.5 & 0.103 & 80 & 1801 & 8999\\
            \bottomrule
        \end{tabular}
    \end{subtable}
    \par \bigskip 
    \begin{subtable}[t]{\textwidth}
        \caption{Rejection rates under alternative parameter configurations}
        \centering
        \begin{tabular}[t]{rrrrrrrrrr}
            \toprule
            $C$ & $\beta_1$ & $\gamma_1$ & $\delta_1$ & $\sigma_1^2$ & $\sigma_2^2$ & Rejection rate & \# Focal & \# Randomizations & \# MC \\
            \midrule
            40 & 0.00 & 0.25 & 0.25 & 0.5 & 0.5 & 0.322 & 80 & 1801 & 7774\\
            40 & 0.00 & 0.50 & 0.50 & 0.5 & 0.5 & 0.652 & 80 & 1801 & 4274\\
            40 & 0.00 & 1.00 & 1.00 & 0.5 & 0.5 & 0.893 & 80 & 1802 & 206\\
            40 & 0.25 & 0.00 & 0.00 & 0.5 & 0.5 & 0.350 & 80 & 1801 & 8999\\
            40 & 0.25 & 0.25 & 0.25 & 0.5 & 0.5 & 0.764 & 80 & 1801 & 7774\\
            40 & 0.25 & 0.50 & 0.50 & 0.5 & 0.5 & 0.933 & 80 & 1801 & 4274\\
            40 & 0.25 & 1.00 & 1.00 & 0.5 & 0.5 & 0.990 & 80 & 1802 & 206\\
            40 & 0.50 & 0.00 & 0.00 & 0.5 & 0.5 & 0.787 & 80 & 1801 & 8999\\
            40 & 0.50 & 0.25 & 0.25 & 0.5 & 0.5 & 0.971 & 80 & 1801 & 7774\\
            40 & 0.50 & 0.50 & 0.50 & 0.5 & 0.5 & 0.995 & 80 & 1801 & 4274\\
            40 & 0.50 & 1.00 & 1.00 & 0.5 & 0.5 & 1.000 & 80 & 1802 & 206\\
            40 & 1.00 & 0.00 & 0.00 & 0.5 & 0.5 & 0.999 & 80 & 1801 & 8999\\
            40 & 1.00 & 0.25 & 0.25 & 0.5 & 0.5 & 1.000 & 80 & 1801 & 7774\\
            40 & 1.00 & 0.50 & 0.50 & 0.5 & 0.5 & 1.000 & 80 & 1801 & 4274\\
            40 & 1.00 & 1.00 & 1.00 & 0.5 & 0.5 & 1.000 & 80 & 1802 & 206\\
            \bottomrule
        \end{tabular}
    \end{subtable}
    \par \bigskip 
    \begin{small}
        Notes: 
        The column ``\# Focal'' reports the number of focal units.
        Rejection rates are computed using only the Monte Carlo replications with the corresponding first-stage selection, and the column ``\# MC'' reports the number of these replications.
        The column ``\# Randomizations'' reports the average number of randomization draws used in the second-stage test across these replications.
    \end{small}
\end{table}

\section{Empirical Illustration} \label{sec:application}

In this section, we illustrate the proposed selective randomization inference using data from the cash transfer experiment in Malawi studied by \citet{baird2011cash}.
In the methodological literature, \citet{baird2018optimal} and \citet{liu2026randomization} use this experiment for their empirical illustrations.
We obtain the data from the World Bank Microdata Library (\citealp{ozler2015schooling}).

\citet{baird2011cash} conducted a randomized experiment under which they assigned unconditional cash transfers (UCTs), conditional cash transfers, and controls to households with school-age girls.
While their goal is to compare the effects of unconditional and conditional cash transfers on schoolgirls' outcomes, we focus on the UCT component of the experiment, following \citet{liu2026randomization}.
Specifically, we assess direct and spillover effects of UCTs on girls enrolled in school at baseline.
The experiment randomized the proportion of baseline schoolgirls offered UCTs within each enumeration area (EA), as well as which girls received offers (see footnote 15 of \citealp{baird2011cash}).
The 42 EAs considered by \citet{liu2026randomization} comprise 27 UCT EAs and 15 EAs in which no baseline schoolgirls were offered cash transfers.
To facilitate the construction of focal subpopulations and the computation of conditional and selective randomization distributions, we restrict attention to the 18 EAs with saturation labels $1/3$ and $2/3$, so that $\mathcal{S} = \{1/3, 2/3\}$.
Nine of these EAs were assigned to each saturation label.
The selected sample comprises 334 baseline schoolgirls, of whom 155 were offered UCTs and outcomes of 322 girls were observed.
Since we restrict our analysis on the subset of the original data of \citet{baird2011cash}, the empirical results reported below should be interpreted with this limitation.

Panel (a) of Table \ref{tab:application} reports the results of the first-stage CRTs of the null hypothesis of no spillover $H_1$.
We consider four binary outcomes: (i) whether a schoolgirl is currently enrolled in school; (ii) whether she reports being able to read a one-page letter in English; (iii) whether she has ever been married; and (iv) whether she has ever been pregnant.
We specify the focal subpopulation by the method described in Section \ref{subsec:first-focal} with setting the number of focal units in each EA with the upper bound in \eqref{eq:bound-K}.
For the choice of test statistic, we employ the Wald test statistic for the linear regression model \eqref{eq:first-statistic}.
None of the four tests rejects $H_1$ at the 10\% significance level.
Thus, we find no statistically significant evidence of spillover effects of UCTs on these outcomes.

Given the first-stage non-rejection, panel (b) of Table \ref{tab:application} reports the results of the second-stage selective randomization inference for direct effects under the assumption of no spillover.
We use units outside the first-stage focal subpopulation with observed outcomes as second-stage focal units, following \eqref{eq:second-accept-focal}.
We employ the absolute difference-in-means statistic in \eqref{eq:second-nonrejection-statistic}.
For three of the four outcomes, our SRTs fail to reject the null hypothesis of no direct effects $\tilde{H}_1$ at the 10\% level.
By contrast, the selective $p$-value for the ever-married outcome is $0.076$, and $\tilde{H}_1$ is rejected at the 10\% level.
Under the additional assumption of homogeneous direct effects, the 90\% selective confidence interval for the average direct effect on the ever-married outcome is $[-0.187, -0.007]$, which excludes zero.

\begin{table}[t]
    \caption{Empirical results} \label{tab:application}
    \begin{subtable}[t]{\textwidth}
        \caption{First-stage conditional randomization tests}
        \centering
        \begin{tabular}[t]{lrrr}
            \toprule
            \multicolumn{1}{c}{Outcome} & \multicolumn{1}{c}{$p$-value} & \multicolumn{1}{c}{\# Focal} & \multicolumn{1}{c}{\# Randomizations} \\
            \midrule
            Currently enrolled & 0.110 & 111 & 1,000,000\\
            English literacy & 0.430 & 111 & 1,000,000\\
            Ever married & 0.511 & 111 & 1,000,000\\
            Ever pregnant & 0.405 & 111 & 1,000,000\\
       \bottomrule
    \end{tabular}
    \end{subtable}
    \par \bigskip 
    \begin{subtable}[t]{\textwidth}
        \caption{Second-stage selective randomization inference after the first-stage non-rejection at $\alpha_1 = 0.1$}
        \centering
        \begin{tabular}[t]{lrcrr}
            \toprule
            \multicolumn{1}{c}{Outcome} & \multicolumn{1}{c}{$p$-value} & \multicolumn{1}{c}{90\% CI} & \multicolumn{1}{c}{\# Focal} & \multicolumn{1}{c}{\# Randomizations} \\
            \midrule
            Currently enrolled & 0.314 & $[-0.044, \phantom{-}0.190]$ & 211 & 904,565\\
            English literacy & 0.266 & $[-0.019, \phantom{-}0.102]$ & 211 & 901,757\\
            Ever married & 0.076 & $[-0.187, -0.007]$ & 211 & 911,322\\
            Ever pregnant  & 0.232 & $[-0.173, \phantom{-}0.027]$ & 211 & 916,393\\
       \bottomrule
    \end{tabular}
    \end{subtable}
    \par \bigskip 
    \begin{small}
        Notes: 
        Panel (a) reports conditional $p$-values for the null hypothesis of no spillover $H_1$ in \eqref{eq:H1}.
        Panel (b) reports selective $p$-values for the null hypothesis of no direct effects $\tilde{H}_1$ in \eqref{eq:tildeH1}, conditional on first-stage non-rejection at the 10\% level.
        Panel (b) also reports 90\% selective confidence intervals for the average direct effects obtained by test inversion under the assumption of homogeneous direct effects.
    \end{small}
\end{table}

\section*{Declaration of the use of generative AI}

During the preparation of this work, the author used ChatGPT for Academic Researchers (ChatGPT-6 Astra) to improve the language and readability of the manuscript and to refine the theoretical and numerical analyses.
After using this tool, the author reviewed and edited the content as necessary and takes full responsibility for the content of the paper.

\section*{Acknowledgements}

This work was supported by JSPS KAKENHI Grant Number 24K04817.

\clearpage

\bibliography{references.bib}

\clearpage

\appendix

\section{Proofs} \label{sec:proof}

\subsection{Proof of Proposition \ref{prop:stage1}}

Fix an arbitrary pair $(\lmathcal{f}_1, \mathcal{Z}_{\lmathcal{f}_1}^1)$ in the joint support of $(\mathcal{F}_1^{\obs}, \mathcal{Z}_{\mathcal{F}_1^{\obs}}^1)$.
Here, $\mathcal{Z}_{\lmathcal{f}_1}^1$ denotes a fixed realization of the imputable set $\mathcal{Z}_{\mathcal{F}_1^{\obs}}^1$ and is not determined by the condition $\mathcal{F}_1^{\obs} = \lmathcal{f}_1$ alone, because $\mathcal{Z}_{\mathcal{F}_1^{\obs}}^1$ depends on the observed assignment $Z^{\obs}$ as well.
More specifically, we take a representative assignment $z^{\obs} \in \mathcal{Z}$ that induces the fixed pair $(\lmathcal{f}_1, \mathcal{Z}_{\lmathcal{f}_1}^1)$. 
Then, we can write
\begin{align} \label{eq:mathcal-Z-f1}
    \mathcal{Z}_{\lmathcal{f}_1}^1 
    = \left\{ z \in \mathcal{Z}: z_i = z_i^{\obs} \text{ for all $i \in \lmathcal{f}_1$} \right\}.
\end{align}
Define $f_{1i} \coloneqq \bm{1}\{ i \in \lmathcal{f}_1 \}$ with the corresponding indicator vector $f_1 \coloneqq ( f_{1i}: i \in [N] )$.
Let $Y_{\lmathcal{f}_1}(z) \coloneqq (Y_i(z): i \in \lmathcal{f}_1)$.

To prove the desired result, it suffices to show that
\begin{align*}
    \Pr \left[ \mathrm{pval}_1(A^{\obs}, Z^{\obs}) \le \alpha_1 \;\middle|\; F_1^{\obs} = f_1, \mathcal{Z}_{\mathcal{F}_1^{\obs}}^1 = \mathcal{Z}_{\lmathcal{f}_1}^1 \right] \le \alpha_1
    \qquad \text{for all $\alpha_1 \in [0, 1]$}.
\end{align*}
Here, conditional on $F_1^{\obs} = f_1$, the event $\mathcal{Z}_{\mathcal{F}_1^{\obs}}^1 = \mathcal{Z}_{\lmathcal{f}_1}^1$ is equivalent to the event $Z^{\obs} \in \mathcal{Z}_{\lmathcal{f}_1}^1$.
Indeed, since $Z^{\obs} \in \mathcal{Z}_{\mathcal{F}_1^{\obs}}^1$ by construction, $\mathcal{Z}_{\mathcal{F}_1^{\obs}}^1 = \mathcal{Z}_{\lmathcal{f}_1}^1$ implies $Z^{\obs} \in \mathcal{Z}_{\lmathcal{f}_1}^1$.
Conversely, if $Z^{\obs} \in \mathcal{Z}_{\lmathcal{f}_1}^1$, then \eqref{eq:mathcal-Z-f1} implies that $Z_i^{\obs} = z_i^{\obs}$ for all $i \in \lmathcal{f}_1$.
Thus, conditional on $F_1^{\obs} = f_1$, \eqref{eq:mathcal-I-F1} gives
\begin{align*}
    \mathcal{Z}_{\mathcal{F}_1^{\obs}}^1
    &= \left\{ z \in \mathcal{Z}: z_i = Z_i^{\obs} \text{ for all $i \in \lmathcal{f}_1$} \right\} \\
    &= \left\{ z \in \mathcal{Z}: z_i = z_i^{\obs} \text{ for all $i \in \lmathcal{f}_1$} \right\} \\
    & = \mathcal{Z}_{\lmathcal{f}_1}^1.
\end{align*}
Accordingly, it suffices to show that
\begin{align*}
    \Pr \left[ \mathrm{pval}_1(A^{\obs}, Z^{\obs}) \le \alpha_1 \;\middle|\; F_1^{\obs} = f_1, Z^{\obs} \in \mathcal{Z}_{\lmathcal{f}_1}^1 \right] \le \alpha_1
    \qquad \text{for all $\alpha_1 \in [0, 1]$}.
\end{align*}

Under $H_1$, we have $Y_{\lmathcal{f}_1}(z) = Y_{\lmathcal{f}_1}(Z^{\obs})$ for all $z \in \mathcal{Z}_{\lmathcal{f}_1}^1$.
Therefore,
\begin{align*}
    \mathrm{pval}_1(A^{\obs}, Z^{\obs})
    = \Pr \Big[ T_1 \big( A, Z, Y_{\lmathcal{f}_1}(Z) \big) \ge T_1 \big( A^{\obs}, Z^{\obs}, Y_{\lmathcal{f}_1}(Z^{\obs}) \big) \; \Big| \; F_1(Z) = f_1, Z \in \mathcal{Z}_{\lmathcal{f}_1}^1 \Big].
\end{align*}
Letting $\mathrm{CDF}_{T_1}(\cdot)$ denote the conditional cumulative distribution function of $-T_1(A, Z, Y_{\lmathcal{f}_1}(Z))$ given $F_1(Z) = f_1$ and $Z \in \mathcal{Z}_{\lmathcal{f}_1}^1$, we can rewrite the conditional $p$-value as
\begin{align*}
    \mathrm{pval}_1(A^{\obs}, Z^{\obs})
    = \mathrm{CDF}_{T_1} \Big( -T_1 \big( A^{\obs}, Z^{\obs}, Y_{\lmathcal{f}_1}(Z^{\obs}) \big) \Big).
\end{align*}
Since $-T_1(A^{\obs}, Z^{\obs}, Y_{\lmathcal{f}_1}(Z^{\obs}))$ also has the conditional cumulative distribution function $\mathrm{CDF}_{T_1}(\cdot)$ given $F_1^{\obs} = f_1$ and $Z^{\obs} \in \mathcal{Z}_{\lmathcal{f}_1}^1$, it holds that
\begin{align*}
    & \Pr \left[ \mathrm{pval}_1(A^{\obs}, Z^{\obs}) \le \alpha_1 \;\middle|\; F_1^{\obs} = f_1, Z^{\obs} \in \mathcal{Z}_{\lmathcal{f}_1}^1 \right] \\
    & = \Pr \left[ \mathrm{CDF}_{T_1} \Big( -T_1 \big( A^{\obs}, Z^{\obs}, Y_{\lmathcal{f}_1}(Z^{\obs}) \big) \Big) \le \alpha_1  \;\middle|\; F_1^{\obs} = f_1, Z^{\obs} \in \mathcal{Z}_{\lmathcal{f}_1}^1  \right] \\
    & \le  \alpha_1,
\end{align*}
where the inequality follows from the fact that $\Pr[ \mathrm{CDF}_R(R) \le \alpha_1 ] \le \alpha_1$ for any real-valued random variable $R$ with cumulative distribution function $\mathrm{CDF}_R$.
\qed 

\subsection{Proof of Theorem \ref{thm:after-rejection}}

Fix an arbitrary $(\lmathcal{f}_1, \lmathcal{f}_2, \mathcal{Z}_{\lmathcal{f}_1}^1, \mathcal{Z}_{\lmathcal{f}_1 \lmathcal{f}_2}^2)$ in the conditional support of $(\mathcal{F}_1^{\obs}, \mathcal{F}_2^{\obs}, \mathcal{Z}_{\mathcal{F}_1^{\obs}}^1, \mathcal{Z}_{\mathcal{F}_1^{\obs} \mathcal{F}_2^{\obs}}^2)$ given $\phi_1^{\obs} = 1$.
Specifically, we fix a tuple $(\lmathcal{f}_1, \lmathcal{f}_2, \mathcal{Z}_{\lmathcal{f}_1}^1, \mathcal{Z}_{\lmathcal{f}_1 \lmathcal{f}_2}^2)$ induced by an arbitrary realization $(a^{\obs}, z^{\obs})$ of $(A^{\obs}, Z^{\obs})$ such that $\phi_1(a^{\obs}, z^{\obs}) = 1$.
Then, we can write
\begin{align} \label{eq:mathcal-Z-f1f2}
\begin{split}
    \mathcal{Z}_{\lmathcal{f}_1 \lmathcal{f}_2}^2
    = \{z \in \mathcal{Z}_{\lmathcal{f}_1}^1:
    & \; E_i(z) \in \{ \mathtt{a}, \mathtt{b} \} \text{ for all $i \in \lmathcal{f}_1 \cup \lmathcal{f}_2$ such that $E_i(z^{\obs}) \in \{ \mathtt{a}, \mathtt{b} \}$,} \\
    & \; E_i(z) = E_i(z^{\obs}) \text{ for all $i \in \lmathcal{f}_1$ such that $E_i(z^{\obs}) \notin \{ \mathtt{a}, \mathtt{b} \}$} \}.
\end{split}
\end{align}
Define $f_{1i} \coloneqq \bm{1}\{ i \in \lmathcal{f}_1 \}$ and $f_{2i} \coloneqq \bm{1}\{ i \in \lmathcal{f}_2 \}$ with the corresponding vectors $f_1 \coloneqq ( f_{1i}: i \in [N] )$ and $f_2 \coloneqq ( f_{2i}: i \in [N] )$.
Let $Y_{\lmathcal{f}_2}(z) \coloneqq (Y_i(z): i \in \lmathcal{f}_2)$.

To prove the desired result, it suffices to show that
\begin{align*}
    \Pr \left[ \mathrm{pval}_2(A^{\obs}, Z^{\obs}) \le \alpha_2 \;\middle|\; \phi_1^{\obs} = 1, F_1^{\obs} = f_1, F_2^{\obs} = f_2, \mathcal{Z}_{\mathcal{F}_1^{\obs} \mathcal{F}_2^{\obs}}^2 = \mathcal{Z}_{\lmathcal{f}_1 \lmathcal{f}_2}^2 \right] \le \alpha_2
\end{align*}
for all $\alpha_2 \in [0, 1]$.
Here, conditional on $F_1^{\obs} = f_1$ and $F_2^{\obs} = f_2$, the event $\mathcal{Z}_{\mathcal{F}_1^{\obs} \mathcal{F}_2^{\obs}}^2 = \mathcal{Z}_{\lmathcal{f}_1 \lmathcal{f}_2}^2$ is equivalent to the event $Z^{\obs} \in \mathcal{Z}_{\lmathcal{f}_1 \lmathcal{f}_2}^2$.
Indeed, since $Z^{\obs} \in \mathcal{Z}_{\mathcal{F}_1^{\obs} \mathcal{F}_2^{\obs}}^2$ by construction, $\mathcal{Z}_{\mathcal{F}_1^{\obs} \mathcal{F}_2^{\obs}}^2 = \mathcal{Z}_{\lmathcal{f}_1 \lmathcal{f}_2}^2$ implies $Z^{\obs} \in \mathcal{Z}_{\lmathcal{f}_1 \lmathcal{f}_2}^2$.
Conversely, if $Z^{\obs} \in \mathcal{Z}_{\lmathcal{f}_1 \lmathcal{f}_2}^2$, then \eqref{eq:mathcal-Z-f1f2} implies that $E_i(Z^{\obs}) \in \{ \mathtt{a}, \mathtt{b} \}$ for all $i \in \lmathcal{f}_1 \cup \lmathcal{f}_2$ such that $E_i(z^{\obs}) \in \{ \mathtt{a}, \mathtt{b} \}$ and that $E_i(Z^{\obs}) = E_i(z^{\obs})$ for all $i \in \lmathcal{f}_1$ such that $E_i(z^{\obs}) \notin \{ \mathtt{a}, \mathtt{b} \}$.
Moreover, since $Z^{\obs} \in \mathcal{Z}_{\lmathcal{f}_1 \lmathcal{f}_2}^2 \subseteq \mathcal{Z}_{\lmathcal{f}_1}^1$, we have $Z_i^{\obs} = z_i^{\obs}$ for all $i \in \lmathcal{f}_1$.
Hence, conditional on $F_1^{\obs} = f_1$, $\mathcal{Z}_{\mathcal{F}_1^{\obs}}^1 = \mathcal{Z}_{\lmathcal{f}_1}^1$.
Thus, conditional on $F_1^{\obs} = f_1$ and $F_2^{\obs} = f_2$, \eqref{eq:mathcal-Z-F1F2} gives
\begin{alignat*}{2}
    \mathcal{Z}_{\mathcal{F}_1^{\obs} \mathcal{F}_2^{\obs}}^2
    & = \{z \in \mathcal{Z}_{\lmathcal{f}_1}^1:
    & & E_i(z) \in \{ \mathtt{a}, \mathtt{b} \} \text{ for all $i \in \lmathcal{f}_1 \cup \lmathcal{f}_2$ such that $E_i(Z^{\obs}) \in \{ \mathtt{a}, \mathtt{b} \}$,} \\
    & & & E_i(z) = E_i(Z^{\obs}) \text{ for all $i \in \lmathcal{f}_1$ such that $E_i(Z^{\obs}) \notin \{ \mathtt{a}, \mathtt{b} \}$} \} \\
    & = \{z \in \mathcal{Z}_{\lmathcal{f}_1}^1:
    & & E_i(z) \in \{ \mathtt{a}, \mathtt{b} \} \text{ for all $i \in \lmathcal{f}_1 \cup \lmathcal{f}_2$ such that $E_i(z^{\obs}) \in \{ \mathtt{a}, \mathtt{b} \}$,} \\
    & & & E_i(z) = E_i(z^{\obs}) \text{ for all $i \in \lmathcal{f}_1$ such that $E_i(z^{\obs}) \notin \{ \mathtt{a}, \mathtt{b} \}$} \} \\
    & = \mathcal{Z}_{\lmathcal{f}_1 \lmathcal{f}_2}^2. &&
\end{alignat*}
Accordingly, it suffices to show that
\begin{align*}
    \Pr \left[ \mathrm{pval}_2(A^{\obs}, Z^{\obs}) \le \alpha_2 \;\middle|\; \phi_1^{\obs} = 1, F_1^{\obs} = f_1, F_2^{\obs} = f_2, Z^{\obs} \in \mathcal{Z}_{\lmathcal{f}_1 \lmathcal{f}_2}^2 \right] \le \alpha_2
\end{align*}
for all $\alpha_2 \in [0, 1]$.

Since the potential outcome vectors $Y_{\lmathcal{f}_1}(z)$ and $Y_{\lmathcal{f}_2}(z)$ are imputable from $Y_{\lmathcal{f}_1}(Z^{\obs})$ and $Y_{\lmathcal{f}_2}(Z^{\obs})$, respectively, for all $z \in \mathcal{Z}_{\lmathcal{f}_1 \lmathcal{f}_2}^2$ under $H_2$ and Assumption \ref{as:correct-exposure}, the first-stage rejection indicator $\phi_1(a, z)$ and the second-stage test statistic $T_2(a, z, Y_{\lmathcal{f}_2}(z))$ are also imputable.
Moreover, since the first-stage focal-subpopulation function satisfies $\mathcal{F}_1(z) = \lmathcal{f}_1$ for all $z \in \mathcal{Z}_{\lmathcal{f}_1}^1$, the first-stage focal subpopulation $\lmathcal{f}_1$ and the corresponding conditional randomization distribution $\Pr(A = \cdot, Z = \cdot \mid F_1(Z) = f_1, Z \in \mathcal{Z}_{\lmathcal{f}_1}^1)$ are unchanged for all $z \in \mathcal{Z}_{\lmathcal{f}_1 \lmathcal{f}_2}^2$.
Hence, $\phi_1(a, z)$ coincides with the rejection indicator obtained by applying the first-stage CRT to the data that would be observed under $(a, z)$.
Then, we have
\begin{align*}
    & \mathrm{pval}_2(A^{\obs}, Z^{\obs}) \\
    & = \Pr \Big[ T_2 \big( A, Z, Y_{\lmathcal{f}_2}(Z) \big) \ge T_2 \big( A^{\obs}, Z^{\obs}, Y_{\lmathcal{f}_2}(Z^{\obs}) \big) \; \Big| \; \phi_1(A, Z) = 1, F_1(Z) = f_1, F_2(Z) = f_2, Z \in \mathcal{Z}_{\lmathcal{f}_1 \lmathcal{f}_2}^2 \Big].
\end{align*}
Letting $\mathrm{CDF}_{T_2}(\cdot)$ denote the conditional cumulative distribution function of $-T_2(A, Z, Y_{\lmathcal{f}_2}(Z))$ given $\phi_1(A, Z) = 1$, $F_1(Z) = f_1$, $F_2(Z) = f_2$, and $Z \in \mathcal{Z}_{\lmathcal{f}_1 \lmathcal{f}_2}^2$, we can rewrite the selective $p$-value as
\begin{align*}
    \mathrm{pval}_2(A^{\obs}, Z^{\obs})
    = \mathrm{CDF}_{T_2} \Big( -T_2 \big( A^{\obs}, Z^{\obs}, Y_{\lmathcal{f}_2}(Z^{\obs}) \big) \Big).
\end{align*}
Since $-T_2(A^{\obs}, Z^{\obs}, Y_{\lmathcal{f}_2}(Z^{\obs}))$ also has the conditional cumulative distribution function $\mathrm{CDF}_{T_2}(\cdot)$ given $\phi_1^{\obs} = 1$, $F_1^{\obs} = f_1$, $F_2^{\obs} = f_2$, and $Z^{\obs} \in \mathcal{Z}_{\lmathcal{f}_1 \lmathcal{f}_2}^2$, it holds that
\begin{align*}
    & \Pr \left[ \mathrm{pval}_2(A^{\obs}, Z^{\obs}) \le \alpha_2 \;\middle|\; \phi_1^{\obs} = 1, F_1^{\obs} = f_1, F_2^{\obs} = f_2, Z^{\obs} \in \mathcal{Z}_{ \lmathcal{f}_1 \lmathcal{f}_2 }^2 \right] \\
    & = \Pr \left[ \mathrm{CDF}_{T_2} \Big( -T_2 \big( A^{\obs}, Z^{\obs}, Y_{\lmathcal{f}_2}(Z^{\obs}) \big) \Big) \le \alpha_2  \;\middle|\; \phi_1^{\obs} = 1, F_1^{\obs} = f_1, F_2^{\obs} = f_2, Z^{\obs} \in \mathcal{Z}_{ \lmathcal{f}_1 \lmathcal{f}_2 }^2 \right] \\
    & \le  \alpha_2,
\end{align*}
by the same argument as in the proof of Proposition \ref{prop:stage1}.
\qed

\subsection{Proof of Theorem \ref{thm:after-accept}}

Using the following three facts, the desired result follows from the same argument as in the proof of Theorem \ref{thm:after-rejection}.
First, since $\tilde{F}_{1i}(z) = 1 - F_{1i}(z)$ for all $i \in [N]$ and $z \in \mathcal{Z}$, conditioning on $\tilde{F}_1^{\obs}$ is equivalent to conditioning on $F_1^{\obs}$.
Second, since $\tilde{\mathcal{Z}}^1 = \mathcal{Z}^1$, the argument in the proof of Proposition \ref{prop:stage1} establishes, conditional on an arbitrary realization of the focal subpopulation $\tilde{\mathcal{F}}_1^{\obs}$, the equivalence between fixing the imputable set and requiring $Z^{\obs}$ to belong to that fixed set.
Third, under Assumptions \ref{as:first-focal} and \ref{as:no-spillover}, the imputability of all potential outcomes under $\tilde{H}_1$ and the restriction $z \in \tilde{\mathcal{Z}}^1$ ensure that the first-stage focal subpopulation $\mathcal{F}_1^{\obs}$ and the conditional randomization distribution $\pi_1$ remain unchanged across any treatment assignment $z \in \tilde{\mathcal{Z}}^1$ and the corresponding saturation assignment $a$.
Hence, $\phi_1(a, z)$ is imputable and coincides with the test result obtained by applying the first-stage test to the data that would be observed under assignment $(a, z)$.
Therefore, the conditional cumulative distribution function argument in the proof of Theorem \ref{thm:after-rejection} applies to the selective $p$-value in \eqref{eq:pval-tildeH1}, with the selection event $\phi_1(A, Z) = 0$ and the test statistic $\tilde{T}_1$, yielding the desired result.
\qed

\section{Computational Details} \label{sec:computation}

\subsection{First-stage conditional randomization test}

To examine the conditional randomization distribution $\pi_1(a, z) = \Pr(A = a, Z = z \mid F_1(Z) = F_1^{\obs}, Z \in \mathcal{Z}^1)$, we introduce the following functions.
Suppose that we construct the focal subpopulation $\mathcal{F}_1^{\obs}$ as described in Section \ref{subsec:first-focal}, implying that $\mathcal{F}_1(z) = \mathcal{F}_1^{\obs}$ for all $z \in \mathcal{Z}$.
For $a = (a_c: c \in [C]) \in \mathcal{A}$, let
\begin{align*}
    q_c(a_c)
    \coloneqq \binom{N_c - K_c}{N_{1c}(a_c) - K_{1c}} \Bigg/ \binom{N_c}{N_{1c}(a_c)},
    \qquad 
    w(a)
    \coloneqq \prod_{c=1}^C q_c(a_c).
\end{align*}
The quantity $q_c(a_c)$ is the probability that, given the saturation assignment $A_c = a_c$, all focal units $i \in \mathcal{F}_1^{\obs}$ in cluster $c$ take the same assignments as the observed assignments $Z_i^{\obs}$.
Correspondingly, the weight $w(a)$ is the probability that, given the saturation assignments $A = a$, all focal units in all clusters take the same assignments as the observed assignments.
Here, we use the following convention for the binomial coefficient: 
\begin{align*}
    \binom{n}{k} = 0
    \qquad
    \text{if $k < 0$ or $k > n$}.
\end{align*}
If $0 \in \mathcal{S}$, this ensures that
\begin{align*}
    q_c(0)
    = \binom{N_c - K_c}{- K_{1c}} \Bigg/ \binom{N_c}{0}
    =
    \begin{cases*}
        1 & \text{if $K_{1c} = 0$,} \\
        0 & \text{if $K_{1c} > 0$.}
    \end{cases*}
\end{align*}
This in turn ensures that the zero saturation label can be assigned only to clusters with $K_{1c} = 0$.
Similarly, if $1 \in \mathcal{S}$,
\begin{align*}
    q_c(1)
    = \binom{N_c - K_c}{N_c - K_{1c}} \Bigg/ \binom{N_c}{N_c}
    =
    \begin{cases*}
        1 & \text{if $K_{0c} = 0$,} \\
        0 & \text{if $K_{0c} > 0$,}
    \end{cases*}
\end{align*}
which ensures that the one saturation label can be assigned only to clusters with $K_{0c} = 0$.

At the end of this section, we show that    
\begin{align} \label{eq:prob-A}
    \Pr \left( A = a \; \middle| \; F_1(Z) = F_1^{\obs}, Z \in \mathcal{Z}^1 \right) 
    = \frac{ w(a) }{ \sum_{a' \in \mathcal{A}} w(a') }
\end{align}
for all $a \in \mathcal{A}$, and that
\begin{align} \label{eq:prob-Z}
\begin{split}
    & \Pr \left( Z = z \; \middle| \; A = a, F_1(Z) = F_1^{\obs}, Z \in \mathcal{Z}^1 \right) \\
    & = \bm{1}\{ z \in \mathrm{supp}(Z \mid A = a) \} \Bigg/ \prod_{c=1}^C \binom{N_c - K_c}{N_{1c}(a_c) - K_{1c}}
\end{split}
\end{align}
for all $a \in \mathcal{A}_{\mathrm{feas}}^{\obs}$ and $z \in \mathcal{Z}^1$, where
\begin{align*}
    \mathcal{A}_{\mathrm{feas}}^{\obs}
    & \coloneqq \{ a \in \mathcal{A}: w(a) > 0 \} \\
    & = \{ a \in \mathcal{A}: K_{1c} \le N_{1c}(a_c) \le N_c - K_{0c} \text{ for all $c \in [C]$} \}.
\end{align*}
In words, $\mathcal{A}_{\mathrm{feas}}^{\obs}$ is the set of feasible saturation assignments in conditional randomization where we fix the treatment assignments of focal units $(Z_i: i \in \mathcal{F}_1^{\obs})$ at the observed treatment assignments $(Z_i^{\obs}: i \in \mathcal{F}_1^{\obs})$.
In particular, $\mathcal{A}_{\mathrm{feas}}^{\obs}$ excludes all saturation assignments under which the zero saturation label is assigned to clusters with $K_{1c} > 0$ or the one saturation label is assigned to clusters with $K_{0c} > 0$.

The results \eqref{eq:prob-A} and \eqref{eq:prob-Z} imply that we can draw $(A, Z)$ from the conditional distribution of $(A, Z)$ given $F_1(Z) = F_1^{\obs}$ and $Z \in \mathcal{Z}^1$ by the following two-step procedure.
First, we draw the vector $A = (A_c : c \in [C])$ from a categorical distribution where each realization $a \in \mathcal{A}$ arises with the probability $w(a) / \sum_{a' \in \mathcal{A}} w(a')$.
Then, conditional on $A$, in each cluster $c$, we choose $N_{1c}(A_c) - K_{1c}$ non-focal units as treated among $N_c - K_c$ non-focal units uniformly without replacement independently across clusters, while the assignments for the focal units are fixed at the observed $Z_i^{\obs}$.
    
When $\mathcal{A}$ is a large set, enumerating all saturation assignments $a \in \mathcal{A}$ and computing their normalized weights $w(a) / \sum_{a' \in \mathcal{A}} w(a')$ can be computationally demanding.
To reduce computational costs, we can adopt the Monte Carlo approximation as follows.
Given the focal subpopulation $\mathcal{F}_1^{\obs}$ and the imputable set $\mathcal{Z}^1$, we begin by specifying large positive integers $R_0$ and $R_1$ such that $R_0 \ge R_1$.
We independently draw $R_0$ candidate vectors $\tilde A^{(1)}, \dots, \tilde A^{(R_0)}$ from the unconditional distribution of $A$.
For each $r_0 \in [R_0]$, we compute the weight $w^{(r_0)} \coloneqq w(\tilde A^{(r_0)})$.
Then, for each $r_1 \in [R_1]$, we conduct the following two steps:
conditional on the candidate vectors, we first draw $\mathtt{index}_{r_1} \in [R_0]$ with probability
\begin{align*}
    \Pr \left( \mathtt{index}_{r_1} = r_0 \; \middle| \; \tilde A^{(1)}, \dots, \tilde A^{(R_0)} \right)
    = \frac{w^{(r_0)}}{\sum_{k=1}^{R_0} w^{(k)}},
    \qquad r_0 \in [R_0],
\end{align*}
independently across repetitions, and set $A^{(r_1)} \coloneqq \tilde A^{(\mathtt{index}_{r_1})}$;
we then choose $N_{1c}(A_c^{(r_1)}) - K_{1c}$ non-focal units as treated among $N_c - K_c$ non-focal units uniformly without replacement, independently across clusters and repetitions conditional on the sampled saturation vectors, while the assignments for the focal units are fixed at the observed $Z_i^{\obs}$.
This procedure leads to the randomized assignment $Z^{(r_1)}$ for $r_1 \in [R_1]$ and approximates the conditional $p$-value by
\begin{align*}
    \hat{p}_1(A^{\obs}, Z^{\obs})
    \coloneqq \frac{1}{1 + R_1} \left[ 1 +  \sum_{r_1 = 1}^{R_1} \bm{1} \left\{ T_1 \left( A^{(r_1)}, Z^{(r_1)}, Y_{\mathcal{F}_1^{\obs}}^{\obs} \right) \ge T_1 \left( A^{\obs}, Z^{\obs}, Y_{\mathcal{F}_1^{\obs}}^{\obs} \right) \right\} \right].
\end{align*}

If $0 \in \mathcal{S}$ and/or $1 \in \mathcal{S}$, then $\sum_{k=1}^{R_0} w^{(k)}$ can be zero, especially when $R_0$ is small and the number of focal units is large relative to the cluster sizes.
In this case, we should substantially increase $R_0$ and regenerate the entire candidate sample.

\subsection{Second-stage selective randomization test after the first-stage rejection}

In general, approximating the selective $p$-value in \eqref{eq:pval-second} by the Monte Carlo method can be computationally demanding because it involves evaluating the second-stage focal-subpopulation selection function $\mathcal{F}_2$ across counterfactual assignments and computing the resulting selective randomization distribution $\pi_2$.
However, if there are only two saturation labels $\mathcal{S} = \mathcal{S}_{\mathrm{int}} = \{ s_1, s_2 \}$ and we set $\mathtt{a} = (1, s_1)$ and $\mathtt{b} = (0, s_1)$, we can approximate the selective $p$-value by an easy-to-implement Monte Carlo procedure.

To be specific, we consider the following setting.
Suppose that we construct the first-stage focal subpopulation $\mathcal{F}_1^{\obs}$ as described in Section \ref{subsec:first-focal} with the number of focal units $K_c$ satisfying $K_c \ge 1$ for any cluster $c$.
For the choice of the first-stage test statistic, we employ the Wald test statistic for testing $\gamma_1 = \delta_1 = 0$ in the linear regression \eqref{eq:first-statistic}.
Similar to Remark \ref{remark:biclique}, we specify the second-stage focal-subpopulation selection function $\mathcal{F}_2$ by
\begin{align*}
    \mathcal{F}_2(z)
    & = \{ i \in [N]: E_i(z) \in \{ \mathtt{a}, \mathtt{b} \} \} \setminus \mathcal{F}_1(z) \\
    & = \{ i \in [N]: A_{c(i)}(z) = s_1 \} \setminus \mathcal{F}_1(z),
\end{align*}
where $A_c(z)$ denotes the saturation assignment of cluster $c$ compatible with the treatment assignment $z$.
In this case, it is easy to see that the imputable set $\mathcal{Z}^2$ defined in \eqref{eq:mathcal-Z-F1F2} reduces to
\begin{align*}
    \mathcal{Z}^2
    = \{z \in \mathcal{Z}^1: A_c(z) = A_c^{\obs} \text{ for all $c \in [C]$} \}.
\end{align*}

We show that, given $\phi_1^{\obs} = 1$, the selective randomization distribution reduces to
\begin{align} \label{eq:pi2-reduction}
    \pi_2(a, z)
    = \Pr( A = a, Z = z \mid A = A^{\obs}, Z \in \mathcal{Z}^1).
\end{align}
For any $z \in \mathcal{Z}^2$, the saturation assignments and the treatment assignments of first-stage focal units remain fixed at their observed values.
Under $H_2$ and Assumption \ref{as:correct-exposure}, their outcomes also remain unchanged.
Hence, the first-stage Wald test statistic is constant across all $z \in \mathcal{Z}^2$.
Since the first-stage focal subpopulation and its treatment assignments remain unchanged, the same first-stage conditional randomization distribution is used when any $z \in \mathcal{Z}^2$ is observed rather than $Z^{\obs}$.
Therefore, we have
\begin{align*}
    \mathrm{pval}_1(A^{\obs}, z)
    = \mathrm{pval}_1(A^{\obs}, Z^{\obs})
    \qquad \text{for all $z \in \mathcal{Z}^2$}.
\end{align*}
Moreover, the second-stage focal-subpopulation selection function $\mathcal{F}_2$ satisfies $\mathcal{F}_2(z) = \mathcal{F}_2^{\obs}$ for all $z \in \mathcal{Z}^2$.
Thus, given $\phi_1^{\obs} = 1$, we obtain \eqref{eq:pi2-reduction}.

To approximate the selective $p$-value in \eqref{eq:pval-second} by the Monte Carlo method, we draw from the selective randomization distribution in \eqref{eq:pi2-reduction} using the following two-step procedure:
(i) we fix $A = A^{\obs}$ and $Z_i = Z_i^{\obs}$ for all first-stage focal units $i \in \mathcal{F}_1^{\obs}$;
(ii) in each cluster $c$, conditional on $A_c = A_c^{\obs}$, we choose $N_{1c}(A_c) - K_{1c}$ non-focal units as treated among $N_c - K_c$ non-focal units uniformly without replacement independently across clusters.
We independently repeat this procedure $R_2$ times to obtain the Monte Carlo sample $\{ (A^{(r_2)}, Z^{(r_2)}): r_2 \in [R_2] \}$.
We then approximate $\mathrm{pval}_2(A^{\obs}, Z^{\obs})$ by
\begin{align*}
    \hat{p}_2(A^{\obs}, Z^{\obs})
    \coloneqq \frac{1}{1 + R_2} \left[ 1 + \sum_{r_2 = 1}^{R_2} \bm{1} \left\{ T_2 \left( A^{(r_2)}, Z^{(r_2)}, Y_{\mathcal{F}_2^{\obs}}^{\obs} \right) \ge T_2 \left( A^{\obs}, Z^{\obs}, Y_{\mathcal{F}_2^{\obs}}^{\obs} \right) \right\} \right].
\end{align*}
This is the Monte Carlo approximation of the selective $p$-value used in the simulation experiments in Section \ref{sec:simulation}.

\subsection{Second-stage selective randomization test after the first-stage non-rejection}

Given the first-stage non-rejection with $\hat{p}_1(A^{\obs}, Z^{\obs}) > \alpha_1$, we approximate the selective $p$-value in \eqref{eq:pval-tildeH1} by the Monte Carlo method as follows.
Let $\mathcal{M}_1 \coloneqq \{ (A^{(r_1)}, Z^{(r_1)}): r_1 \in [R_1] \}$ be the first-stage Monte Carlo sample.
Under the null hypothesis of no direct effects $\tilde{H}_1$ and Assumption \ref{as:no-spillover}, we can reconstruct the counterfactual first-stage test results for each $(a, z) \in \mathcal{M}_1$ by
\begin{align*}
    \hat{\phi}_1(a, z)
    \coloneqq \bm{1} \left\{ \hat{p}_1(a, z) \le \alpha_1 \right \},
\end{align*}
where
\begin{align*}
    \hat{p}_1(a, z)
    \coloneqq \frac{1}{1 + R_1} \left[ 1 + \sum_{k=1}^{R_1} \bm{1}\left\{ T_1 \left( A^{(k)}, Z^{(k)}, Y_{\mathcal{F}_1^{\obs}}^{\obs} \right) \ge T_1 \left( a, z, Y_{\mathcal{F}_1^{\obs}}^{\obs} \right) \right\} \right].
\end{align*}
Define
\begin{align*}
    \tilde{\mathcal{R}}_1
    \coloneqq \left\{ r_1 \in [R_1]: \hat{\phi}_1(A^{(r_1)}, Z^{(r_1)}) = 0 \right\},
\end{align*}
which is the set of indexes $r_1 \in [R_1]$ such that the corresponding saturation and treatment assignments $(A^{(r_1)}, Z^{(r_1)})$ lead to the counterfactual first-stage non-rejection $\hat{\phi}_1(A^{(r_1)}, Z^{(r_1)}) = 0$.
Then, the selective $p$-value in \eqref{eq:pval-tildeH1} can be approximated by
\begin{align*}
    \hat{\tilde{p}}_1(A^{\obs}, Z^{\obs})
    \coloneqq \frac{1}{1 + |\tilde{\mathcal{R}}_1|} \left[ 1 + \sum_{\tilde{r}_1 \in \tilde{\mathcal{R}}_1} \bm{1} \left\{ \tilde{T}_1 \left( A^{(\tilde{r}_1)}, Z^{(\tilde{r}_1)}, Y_{\tilde{\mathcal{F}}_1^{\obs}}^{\obs} \right) \ge \tilde{T}_1 \left( A^{\obs}, Z^{\obs}, Y_{\tilde{\mathcal{F}}_1^{\obs}}^{\obs} \right) \right\} \right].
\end{align*}
This is the Monte Carlo approximation of the selective $p$-value used in the simulation experiments and the empirical illustration in Sections \ref{sec:simulation} and \ref{sec:application}, respectively.

\subsection{Proofs of (\ref{eq:prob-A}) and (\ref{eq:prob-Z})}

Recall that in each cluster $c$, conditional on $A_c = a_c$, there are $\binom{N_c}{N_{1c}(a_c)}$ patterns to assign treatments to its members.
Among them, $\binom{ N_c - K_c }{ N_{1c}(a_c) - K_{1c} }$ patterns keep the assignments of $K_c$ focal units in this cluster fixed at the observed $Z_i^{\obs}$.
Thus, for any $a \in \mathcal{A}$, we have
\begin{align*}
    \Pr \left( Z \in \mathcal{Z}^1 \; \middle| \; A = a \right)
    & = \prod_{c = 1}^C \left[ \binom{N_c - K_c}{N_{1c}(a_c) - K_{1c}} \Bigg/ \binom{N_c}{N_{1c}(a_c)} \right] \\
    & = \prod_{c = 1}^C q_c(a_c),
\end{align*}
implying that
\begin{align*}
    \Pr \left( Z \in \mathcal{Z}^1 \right)
    & = \sum_{a' \in \mathcal{A}} \Pr( A = a' ) \Pr \left( Z \in \mathcal{Z}^1 \; \middle| \; A = a' \right) \\
    & = \sum_{a' \in \mathcal{A}} \frac{1}{|\mathcal{A}|} \prod_{c=1}^C q_c(a'_c).
\end{align*}
Therefore, for any $a \in \mathcal{A}$,
\begin{align*}
    \Pr \left( A = a \; \middle| \; F_1(Z) = F_1^{\obs}, Z \in \mathcal{Z}^1 \right) 
    & = \Pr \left( A = a \; \middle| \; Z \in \mathcal{Z}^1 \right) \\
    & = \frac{ \Pr( A = a, Z \in \mathcal{Z}^1 ) }{ \Pr( Z \in \mathcal{Z}^1 ) } \\
    & = \frac{ \prod_{c=1}^C q_c(a_c) }{ \sum_{a' \in \mathcal{A}} \prod_{c=1}^C q_c(a'_c) },
\end{align*}
which proves \eqref{eq:prob-A}.
Meanwhile, for any $a \in \mathcal{A}_{\mathrm{feas}}^{\mathrm{obs}}$, conditional on $A = a$ and $Z \in \mathcal{Z}^1$, we select $N_{1c}(a_c) - K_{1c}$ of the $N_c - K_c$ non-focal units in each cluster $c$ uniformly without replacement to receive treatment, independently across clusters.
This implies that 
\begin{align*}
    \Pr \left( Z = z \; \middle| \; A = a, F_1(Z) = F_1^{\obs}, Z \in \mathcal{Z}^1 \right)
    = \bm{1}\{ z \in \mathrm{supp}(Z \mid A = a) \} \Bigg/ \prod_{c=1}^C \binom{N_c - K_c}{N_{1c}(a_c) - K_{1c}}
\end{align*}
for any $z \in \mathcal{Z}^1$.
This proves \eqref{eq:prob-Z}.
\qed

\section{Additional Simulation Results} \label{sec:simulation2}

To highlight the importance of conditioning on the result of a pretest for interference, we conduct additional Monte Carlo experiments to examine two conventional tests: 
a CRT for the null hypothesis of zero exposure contrast $H_2$ with $\mathtt{a} = (1, 1/3)$ and $\mathtt{b} = (0, 1/3)$, and the canonical FRT.
Neither test conditions on the first-stage test result.
For the CRT for $H_2$, we specify the focal subpopulation and the corresponding imputable set as follows:
\begin{align*}
    \mathcal{F}_{\{ \mathtt{a}, \mathtt{b} \}} 
    & \coloneqq \{ i \in [N]: E_i^{\obs} \in \{ \mathtt{a}, \mathtt{b} \} \}, \\
    \mathcal{Z}_{\{ \mathtt{a}, \mathtt{b} \}} 
    & \coloneqq \{ z \in \mathcal{Z}: E_i(z) \in \{ \mathtt{a}, \mathtt{b} \} \text{ for all $i \in \mathcal{F}_{\{ \mathtt{a}, \mathtt{b} \}}$} \}.
\end{align*}
The canonical FRT tests Fisher's sharp null of no treatment effects using all units and the unconditional randomization distribution of $(A, Z)$ under the randomized saturation design.
We use the absolute difference-in-means test statistics in \eqref{eq:second-rejection-statistic} and \eqref{eq:second-nonrejection-statistic} while replacing $\mathcal{F}_2^{\obs}$ with $\mathcal{F}_{\{ \mathtt{a}, \mathtt{b} \}}$ and $\tilde{\mathcal{F}}_1^{\obs}$ with $[N]$, respectively.
We employ the same simulation design as in Section \ref{sec:simulation}.

Tables \ref{tab:simulationCRT} and \ref{tab:simulationFRT} report the empirical Type I error rates for the conventional CRT and the canonical FRT, respectively, at the 10\% level for the parameter configurations $\beta_1 = \gamma_1 = \delta_1 = 0$, $\sigma_1^2 \in \{ 0.01, 0.1, 0.5, 0.9, 0.99 \}$, and $\sigma_2^2 = 1 - \sigma_1^2$.
We consider these choices of $\sigma_1^2$ and $\sigma_2^2$ to examine how the variances of the unit-level and cluster-level error terms affect the Type I error rates.
Panel (a) of Table \ref{tab:simulationCRT} reports the empirical Type I error rates conditional on the first-stage rejection, whereas panel (a) of Table \ref{tab:simulationFRT} reports those conditional on the first-stage non-rejection.
Panel (b) of each table reports the empirical Type I error rates across all Monte Carlo replications.

While both tests have empirical Type I error rates close to the nominal level of 10\% across all Monte Carlo replications, their rejection rates conditional on the first-stage test result can deviate substantially from the nominal level.
When $\sigma_1^2 = 0.9$ and $\sigma_1^2 = 0.99$, the conventional CRT has empirical selective Type I error rates of 16.8\% and 16.1\%, respectively, after the first-stage rejection, exceeding 1.5 times the nominal level.
By contrast, when $\sigma_1^2 = 0.01$ and $\sigma_1^2 = 0.1$, the canonical FRT is conservative after the first-stage non-rejection, with empirical selective Type I error rates of 6.9\% and 8.0\%, respectively.
These simulation results emphasize the importance of selective randomization inference that explicitly conditions on the result of the first-stage test for interference.

\section{Algorithms} \label{sec:algoirthm}

This appendix presents Algorithms \ref{alg:first-stage}, \ref{alg:second-stage-rejection}, and \ref{alg:second-stage-non-rejection} to summarize the proposed testing procedures.
These algorithms describe the exact testing procedures under the conditions of Proposition \ref{prop:stage1} and Theorems \ref{thm:after-rejection} and \ref{thm:after-accept}.
Monte Carlo implementations are discussed in Appendix \ref{sec:computation}.

\begin{table}[!p]
    \caption{Simulation results for the conventional CRT for $H_2$ with $\mathtt{a} = (1, 1/3)$ and $\mathtt{b} = (0, 1/3)$ at the 10\% level} \label{tab:simulationCRT}
    \begin{subtable}[t]{\textwidth}
        \caption{Type I error rates after the first-stage rejection}
        \centering
        \begin{tabular}[t]{rrrrrrrrrr}
            \toprule
            $C$ & $\beta_1$ & $\gamma_1$ & $\delta_1$ & $\sigma_1^2$ & $\sigma_2^2$ & Rejection rate & \# Focal & \# Randomizations & \# MC \\
            \midrule
            40 & 0 & 0 & 0 & 0.01 & 0.99 & 0.102 & 60 & 2000 & 1026\\
            40 & 0 & 0 & 0 & 0.10 & 0.90 & 0.105 & 60 & 2000 & 1036\\
            40 & 0 & 0 & 0 & 0.50 & 0.50 & 0.138 & 60 & 2000 & 1001\\
            40 & 0 & 0 & 0 & 0.90 & 0.10 & 0.168 & 60 & 2000 & 1015\\
            40 & 0 & 0 & 0 & 0.99 & 0.01 & 0.161 & 60 & 2000 & 1021\\
            \bottomrule
        \end{tabular}
    \end{subtable}
    \par \bigskip 
    \begin{subtable}[t]{\textwidth}
        \caption{Type I error rates for all replications}
        \centering
        \begin{tabular}[t]{rrrrrrrrrr}
            \toprule
            $C$ & $\beta_1$ & $\gamma_1$ & $\delta_1$ & $\sigma_1^2$ & $\sigma_2^2$ & Rejection rate & \# Focal & \# Randomizations & \# MC \\
            \midrule
            40 & 0 & 0 & 0 & 0.01 & 0.99 & 0.104 & 60 & 2000 & 10000\\
            40 & 0 & 0 & 0 & 0.10 & 0.90 & 0.104 & 60 & 2000 & 10000\\
            40 & 0 & 0 & 0 & 0.50 & 0.50 & 0.104 & 60 & 2000 & 10000\\
            40 & 0 & 0 & 0 & 0.90 & 0.10 & 0.104 & 60 & 2000 & 10000\\
            40 & 0 & 0 & 0 & 0.99 & 0.01 & 0.104 & 60 & 2000 & 10000\\
            \bottomrule
        \end{tabular}
    \end{subtable}
    \par \bigskip 
    \begin{small}
        Notes: 
        The column ``\# Focal'' reports the number of focal units.
        The column ``\# MC'' reports the number of Monte Carlo replications.
        The column ``\# Randomizations'' reports the number of randomization draws.
    \end{small}
\end{table}

\begin{table}[!p]
    \caption{Simulation results for the canonical FRT at the 10\% level} \label{tab:simulationFRT}
    \begin{subtable}[t]{\textwidth}
        \caption{Type I error rates after the first-stage non-rejection}
        \centering
        \begin{tabular}[t]{rrrrrrrrrr}
            \toprule
            $C$ & $\beta_1$ & $\gamma_1$ & $\delta_1$ & $\sigma_1^2$ & $\sigma_2^2$ & Rejection rate & \# Focal & \# Randomizations & \# MC \\
            \midrule
            40 & 0 & 0 & 0 & 0.01 & 0.99 & 0.069 & 120 & 2000 & 8974\\
            40 & 0 & 0 & 0 & 0.10 & 0.90 & 0.080 & 120 & 2000 & 8964\\
            40 & 0 & 0 & 0 & 0.50 & 0.50 & 0.101 & 120 & 2000 & 8999\\
            40 & 0 & 0 & 0 & 0.90 & 0.10 & 0.103 & 120 & 2000 & 8985\\
            40 & 0 & 0 & 0 & 0.99 & 0.01 & 0.101 & 120 & 2000 & 8979\\
            \bottomrule
        \end{tabular}
    \end{subtable}
    \par \bigskip 
    \begin{subtable}[t]{\textwidth}
        \caption{Type I error rates for all replications}
        \centering
        \begin{tabular}[t]{rrrrrrrrrr}
            \toprule
            $C$ & $\beta_1$ & $\gamma_1$ & $\delta_1$ & $\sigma_1^2$ & $\sigma_2^2$ & Rejection rate & \# Focal & \# Randomizations & \# MC \\
            \midrule
            40 & 0 & 0 & 0 & 0.01 & 0.99 & 0.102 & 120 & 2000 & 10000\\
            40 & 0 & 0 & 0 & 0.10 & 0.90 & 0.103 & 120 & 2000 & 10000\\
            40 & 0 & 0 & 0 & 0.50 & 0.50 & 0.104 & 120 & 2000 & 10000\\
            40 & 0 & 0 & 0 & 0.90 & 0.10 & 0.103 & 120 & 2000 & 10000\\
            40 & 0 & 0 & 0 & 0.99 & 0.01 & 0.101 & 120 & 2000 & 10000\\
            \bottomrule
        \end{tabular}
    \end{subtable}
    \par \bigskip 
    \begin{small}
        Notes: 
        The column ``\# Focal'' reports the number of focal units.
        The column ``\# MC'' reports the number of Monte Carlo replications.
        The column ``\# Randomizations'' reports the number of randomization draws.
    \end{small}
\end{table}

\clearpage 

\begin{algorithm}[!p]
    \caption{Conditional randomization test for the null hypothesis of no spillover $H_1$}
    \label{alg:first-stage}
    \begin{spacing}{1.1}
    \noindent\textbf{Input:} The observed data $(A^{\obs}, Z^{\obs}, Y^{\obs})$, the randomized saturation design, a pre-specified focal-subpopulation selection function $\mathcal{F}_1$, a test statistic $T_1$, and a significance level $\alpha_1$.
    If the selection function uses auxiliary randomness, hold its observed realization fixed throughout.
    \begin{enumerate}
        \item Select the focal subpopulation $\mathcal{F}_1^{\obs} = \mathcal{F}_1(Z^{\obs})$ and compute its membership indicator vector $F_1^{\obs}$.
        Construct the imputable set:
        \begin{align*}
            \mathcal{Z}^1
            = \{ z \in \mathcal{Z}: z_i = Z_i^{\obs} \text{ for all } i \in \mathcal{F}_1^{\obs} \}.
        \end{align*}
        \item Construct the conditional randomization distribution:
        \begin{align*}
            \pi_1(a, z)
            = \Pr \left( A = a, Z = z \; \middle| \; F_1(Z) = F_1^{\obs}, Z \in \mathcal{Z}^1 \right).
        \end{align*}
        \item Compute the observed test statistic $T_1(A^{\obs}, Z^{\obs}, Y_{\mathcal{F}_1^{\obs}}^{\obs})$.
        \item Compute the conditional $p$-value:
        \begin{align*}
            \mathrm{pval}_1(A^{\obs}, Z^{\obs})
             = \sum_{(a, z) \in \mathrm{supp}(\pi_1)} 
            \bm{1} \left\{ T_1 \left( a, z, Y_{\mathcal{F}_1^{\obs}}^{\obs} \right) \ge T_1 \left( A^{\obs}, Z^{\obs}, Y_{\mathcal{F}_1^{\obs}}^{\obs} \right) \right\} \pi_1(a, z).
        \end{align*}
        \item Reject $H_1$ if $\mathrm{pval}_1(A^{\obs}, Z^{\obs}) \le \alpha_1$.
    \end{enumerate}
    \noindent\textbf{Output:} The focal subpopulation $\mathcal{F}_1^{\obs}$, the imputable set $\mathcal{Z}^1$, the conditional randomization distribution $\pi_1$, the conditional $p$-value $\mathrm{pval}_1(A^{\obs}, Z^{\obs})$, and the test result $\phi_1^{\obs} \coloneqq \bm{1}\{ \mathrm{pval}_1(A^{\obs}, Z^{\obs}) \le \alpha_1 \}$.
    \end{spacing}
\end{algorithm}

\begin{algorithm}[!p]
    \caption{Selective randomization test for the null hypothesis of zero exposure contrast $H_2$ under Assumption \ref{as:correct-exposure} after the first-stage rejection}
    \label{alg:second-stage-rejection}
    \begin{spacing}{1.1}
    \noindent\textbf{Input:} The observed data $(A^{\obs}, Z^{\obs}, Y^{\obs})$, the randomized saturation design, the first-stage inputs and outputs from Algorithm \ref{alg:first-stage} with $\phi_1^{\obs} = 1$, the exposure mapping $E_i(Z) = (Z_i, A_{c(i)})$ satisfying Assumption \ref{as:correct-exposure}, pre-specified exposure values $\mathtt{a}$ and $\mathtt{b}$, a pre-specified focal-subpopulation selection function $\mathcal{F}_2$, a test statistic $T_2$, and a significance level $\alpha_2$.
    If the focal-subpopulation selection functions $\mathcal{F}_1$ and $\mathcal{F}_2$ use auxiliary randomness, hold its observed realization fixed throughout.
    \begin{enumerate}
        \item Select the focal subpopulation $\mathcal{F}_2^{\obs} = \mathcal{F}_2(Z^{\obs})$ and compute its membership indicator vector $F_2^{\obs}$.
        With $\mathcal{Z}_i^2$ defined in \eqref{eq:mathcal-Z-2}, construct the imputable set:
        \begin{align*}
            \mathcal{Z}^2
            = \mathcal{Z}^1 \cap \left( \bigcap_{i \in \mathcal{F}_1^{\obs} \cup \mathcal{F}_2^{\obs}} \mathcal{Z}_i^2 \right).
        \end{align*}
        \item For each $z \in \mathcal{Z}^2$ and compatible saturation assignment $a$, impute the first-stage rejection indicator $\phi_1(a, z)$ under $H_2$ using $Y_{\mathcal{F}_1^{\obs}}(z) = Y_{\mathcal{F}_1^{\obs}}^{\obs}$, the test statistic $T_1$, the level $\alpha_1$, and the original conditional randomization distribution $\pi_1$ from Algorithm \ref{alg:first-stage}.
        \item Construct the selective randomization distribution:
        \begin{align*}
            \pi_2(a, z)
            = \Pr \Big(A = a, Z = z \; \Big| \; \phi_1(A, Z) = 1, F_1(Z) = F_1^{\obs}, F_2(Z) = F_2^{\obs}, Z \in \mathcal{Z}^2 \Big).
        \end{align*}
        \item Compute the observed test statistic $T_2(A^{\obs}, Z^{\obs}, Y_{\mathcal{F}_2^{\obs}}^{\obs})$.
        \item Compute the selective $p$-value:
        \begin{align*}
            \mathrm{pval}_2(A^{\obs}, Z^{\obs})
            = \sum_{(a, z) \in \mathrm{supp}(\pi_2)}
            \bm{1} \left\{ T_2 \left( a, z, Y_{\mathcal{F}_2^{\obs}}^{\obs} \right) \ge T_2 \left( A^{\obs}, Z^{\obs}, Y_{\mathcal{F}_2^{\obs}}^{\obs} \right) \right\} \pi_2(a, z).
        \end{align*}
        \item Reject $H_2$ if $\mathrm{pval}_2(A^{\obs}, Z^{\obs}) \le \alpha_2$.
    \end{enumerate}
    \noindent\textbf{Output:} The focal subpopulation $\mathcal{F}_2^{\obs}$, the imputable set $\mathcal{Z}^2$, the selective $p$-value $\mathrm{pval}_2(A^{\obs}, Z^{\obs})$, and the test result $\phi_2^{\obs} \coloneqq \bm{1}\{ \mathrm{pval}_2(A^{\obs}, Z^{\obs}) \le \alpha_2 \}$.
    \end{spacing}
\end{algorithm}

\begin{algorithm}[!p]
    \caption{Selective randomization test for the null hypothesis of no direct effects $\tilde{H}_1$ under Assumption \ref{as:no-spillover} after the first-stage non-rejection}
    \label{alg:second-stage-non-rejection}
    \begin{spacing}{1.1}
    \noindent\textbf{Input:} The observed data $(A^{\obs}, Z^{\obs}, Y^{\obs})$, the randomized saturation design, the first-stage inputs and outputs from Algorithm \ref{alg:first-stage} with $\phi_1^{\obs} = 0$, a test statistic $\tilde{T}_1$, and a significance level $\tilde{\alpha}_1$.
    If the first-stage focal-subpopulation selection function $\mathcal{F}_1$ uses auxiliary randomness, hold its observed realization fixed throughout.
    \begin{enumerate}
        \item Define $\tilde{\mathcal{F}}_1(z) = [N] \setminus \mathcal{F}_1(z)$ for all $z \in \mathcal{Z}$.
        Set $\tilde{\mathcal{F}}_1^{\obs} = \tilde{\mathcal{F}}_1(Z^{\obs})$ and compute its membership indicator vector $\tilde{F}_1^{\obs}$.
        Construct the imputable set:
        \begin{align*}
            \tilde{\mathcal{Z}}^1
            = \mathcal{Z}^1
            = \{ z \in \mathcal{Z}: z_i = Z_i^{\obs} \text{ for all } i \in \mathcal{F}_1^{\obs} \}.
        \end{align*}
        \item For each $z \in \tilde{\mathcal{Z}}^1$ and compatible saturation assignment $a$, impute the first-stage rejection indicator $\phi_1(a, z)$ under $\tilde{H}_1$ and Assumption \ref{as:no-spillover} using $Y_{\mathcal{F}_1^{\obs}}(z) = Y_{\mathcal{F}_1^{\obs}}^{\obs}$, the test statistic $T_1$, the level $\alpha_1$, and the original conditional randomization distribution $\pi_1$ from Algorithm~\ref{alg:first-stage}.
        \item Construct the selective randomization distribution:
        \begin{align*}
            \tilde{\pi}_1(a, z)
            = \Pr \Big(A = a, Z = z \; \Big| \; \phi_1(A, Z) = 0, \tilde{F}_1(Z) = \tilde{F}_1^{\obs}, Z \in \tilde{\mathcal{Z}}^1 \Big).
        \end{align*}
        \item Compute the observed test statistic $\tilde{T}_1(A^{\obs}, Z^{\obs}, Y_{\tilde{\mathcal{F}}_1^{\obs}}^{\obs})$.
        \item Compute the selective $p$-value:
        \begin{align*}
            \tilde{\mathrm{pval}}_1(A^{\obs}, Z^{\obs})
            = \sum_{(a, z) \in \mathrm{supp}(\tilde{\pi}_1)}
            \bm{1} \left\{ \tilde{T}_1 \left( a, z, Y_{\tilde{\mathcal{F}}_1^{\obs}}^{\obs} \right) \ge \tilde{T}_1 \left( A^{\obs}, Z^{\obs}, Y_{\tilde{\mathcal{F}}_1^{\obs}}^{\obs} \right) \right\} \tilde{\pi}_1(a, z).
        \end{align*}
        \item Reject $\tilde{H}_1$ if $\tilde{\mathrm{pval}}_1(A^{\obs}, Z^{\obs}) \le \tilde{\alpha}_1$.
    \end{enumerate}
    \noindent\textbf{Output:} The focal subpopulation $\tilde{\mathcal{F}}_1^{\obs}$, the imputable set $\tilde{\mathcal{Z}}^1$, the selective $p$-value $\tilde{\mathrm{pval}}_1(A^{\obs}, Z^{\obs})$, and the test result $\tilde{\phi}_1^{\obs} \coloneqq \bm{1}\{ \tilde{\mathrm{pval}}_1(A^{\obs}, Z^{\obs}) \le \tilde{\alpha}_1 \}$.
    \end{spacing}
\end{algorithm}

\end{document}